\documentclass[aps,preprintnumbers,twocolumn,amsmath,amssymb,floatfix,superscriptaddress,pra]{revtex4-1}
\pdfoutput=1
\usepackage[utf8]{inputenc}
\usepackage{amsmath}
\usepackage{amsfonts}
\usepackage{braket}
\usepackage{tensor}
\usepackage{amssymb}
\usepackage{graphicx}
\usepackage{titlesec}
\usepackage[titletoc,toc,title]{appendix}
\usepackage[colorlinks=true,linkcolor=blue, citecolor=blue, urlcolor=blue, bookmarks]{hyperref}

\def\be{\begin{equation}}
\def\ee{\end{equation}}
\def\bea{\begin{eqnarray}}
\def\eea{\end{eqnarray}}

\def\XXint#1#2#3{{\setbox0=\hbox{$#1{#2#3}{\int}$}
         \vcenter{\hbox{$#2#3$}}\kern-.5\wd0}}

\begin{document}

\title{Quench action and large deviations: work statistics in the one-dimensional Bose gas}

\author{Gabriele Perfetto}
\affiliation{SISSA - International School for Advanced Studies, via Bonomea 265, 34136 Trieste, Italy.}
\affiliation{INFN, Sezione di Trieste, via Bonomea 265, 34136, Trieste, Italy.}
\author{Lorenzo Piroli}
\affiliation{Max-Planck-Institut f\"ur Quantenoptik, Hans-Kopfermann-Str.~1, 85748 Garching, Germany.}
\author{Andrea Gambassi}
\affiliation{SISSA - International School for Advanced Studies, via Bonomea 265, 34136 Trieste, Italy.}
\affiliation{INFN, Sezione di Trieste, via Bonomea 265, 34136, Trieste, Italy.}
\begin{abstract}

We study the statistics of large deviations of the intensive work done in an interaction quench of a one-dimensional Bose gas with a large number $N$ of particles, system size $L$ and fixed density. We consider the case in which the system is initially prepared in the non-interacting ground state and a repulsive interaction is suddenly turned on. For large deviations of the work below its mean value, we show that the large deviation principle holds by means of the quench action approach. Using the latter, we compute exactly the so-called rate function, and study its properties analytically. In particular, we find that fluctuations close to the mean value of the work exhibit a marked non-Gaussian behavior, even though their probability is always exponentially suppressed below it as $L$ increases. Deviations larger than the mean value, instead, exhibit an algebraic decay, whose exponent can not be determined directly by large-deviation theory. Exploiting the exact Bethe ansatz representation of the eigenstates of the Hamiltonian, we calculate this exponent for vanishing particle density. Our approach can be straightforwardly generalized to quantum quenches in other interacting integrable systems.

\end{abstract}

\maketitle

\section{Introduction}
\label{sec:intro}

A defining feature of the quantum theory is the presence of statistical fluctuations in the outcome of the measurement of any physical observable: given a collection of systems prepared in the same quantum state, the result of identical measurements will be generally different, leading to a non-trivial probability distribution function for the outcomes. For many-body systems, it is often very difficult to determine the latter and to give predictions beyond the corresponding mean value. This is especially true out of equilibrium, where exceptional computational challenges arise, even in prototypical solvable models \cite{CaEM16}. On the other hand, old atomic experiments have recently proven that the full probability distribution of certain observables can be probed in mesoscopic systems \cite{HLSI08,AJKB10,JABK11,KPIS10,KISD11}, motivating new efforts in the theoretical study of the fluctuations of quantum measurements in many-body physics, both in \cite{ChDe07,LaFe08,ia-13,sk-13,e-13,k-14,mcsc-15,sp-17,CoEG17,nr-17,hb-17} and out \cite{EiRW03,gadp-06,er-13,lddz-15,GrEC18,bpc-18,BaPi18,CoEs19} of equilibrium.

In this respect, a natural quantity which can be investigated is the work done upon changing some of the system's parameters \cite{Jarz97,TalLutz07,Silv08,GooldPlastina18}, as, e.g., in a quantum quench \cite{cc-06,PolkSeng11}. Crucially, its statistics has now proven to encode important information on the internal dynamics of the system \cite{Silv08,PaSi09,SmSi13,MaSi14,PaSo14,Palm15,Rotondo18} and to display interesting features such as, most prominently, an emergent universal behavior \cite{BDKP11}, in quenches near a critical point \cite{GaSi12,GaSi11,GaSi11,GaSi12,SoGS13}. Furthermore, the statistics of the work is a valuable tool for studying dynamical phase transitions \cite{HePK13, Heyl14,Heyl18} and for detecting them \cite{HePK13, Heyl14,AbKe16}. 

Despite its importance, the explicit calculation of the statistics of the work remain a difficult task even in simplified protocols such as quenches \cite{cc-06,PolkSeng11} and analytic results in the presence of interactions exist only in a few special cases~\cite{PaSo14,Palm15, RyAn18}. Yet, some of its features can be understood based on general arguments and on the analysis of non-interacting models \cite{Silv08}. For instance, several studies have shown the existence of a universal edge singularity at the lowest threshold of the probability distribution $P(W)$ of the extensive work $W$ \cite{Silv08,SoGS13}, and its robustness against different non-equilibrium protocols \cite{SmSi13}. These features have been also verified in integrable quantum field theories \cite{PaSo14,Palm15,RyAn18}, by exploiting the fact that $P(W)$ can be obtained as the Fourier transform of the so-called post-quench Loschmidt amplitude \cite{Silv08,HePK13}.

In addition to $W$, one could also study the probability distribution of the intensive work $w=W/L$, where $L$ is the system size. As $W$ is an extensive variable, one generically expects that both its mean value $\langle W \rangle$ and its fluctuations $\langle (\Delta W)^2 \rangle = \langle W^2 \rangle - \langle W \rangle^2$ grow proportionally to $L$ as $L$ increases, the latter statement being true whenever $W$ can be seen as resulting from the sum of a number proportional to $L$ of almost independent contributions. This implies that the typical fluctuations in the value of $W$ are of order $\sqrt{L}$, i.e., those in $w$, of order $1/\sqrt{L}$, vanish as $L \rightarrow \infty$. Correspondingly, the distribution function $p(w)$ of $w$ approaches a delta function $\delta(w-\bar{w})$ which selects the average value $\bar{w} = \langle W \rangle/L$. On the other hand, fluctuations of the value of $w$ away from $\bar{w}$ corresponds to fluctuations of order $L$ in $W$, i.e., to large and atypical fluctuations, which are increasingly rare as $L$ grows. For free Bosonic and Fermionic models it was found \cite{GaSi12} that $p(w) \sim {\rm exp}[-L I(w)]$ where $I(w)$ is the so-called rate function of large deviation theory \cite{Touc09}, a non-negative function which vanishes for $w=\bar{w}$ and which controls the rate of exponential suppression of large deviations. Importantly, it was shown in Ref.~\cite{GaSi12} that $I(w)$ provides insight into the universal properties of the system for $w\ll \bar{w}$. Furthermore, it was argued that its qualitative behavior can be inferred based on the knowledge of a few parameters of the quench. The analysis of Ref.~\cite{GaSi12} also revealed that, for free bosonic models starting from a critical initial state,  a further universal behavior appears in the regime $w>\bar{w}$, where $p(w)$ displays a transition from the aforementioned exponential decay to an algebraic decay. This transition is analogous to well-known phenomenon of Bose-Einstein condensation in quantum statistical mechanics \cite{Huang}.

In the presence of interactions, the picture presented in Ref.~\cite{GaSi12} remains qualitatively correct but the rate function $I(w)$ is very hard to compute in practice. However, accounting for interactions is obviously important, for instance for quantitative comparisons between theoretical predictions and experiments, in which the interactions are usually non-negligible. In this paper, we show that the statistics of the large deviations of the work done during a global quench can be determined quite generically for any interacting integrable system by means of the recently introduced quench action method \cite{CaEs13,DWBC14,Caux16}. In order to exemplify this approach, we study a prototypical example, namely an interaction quench in the Lieb-Liniger model \cite{LiLi63,Lieb63} describing a one-dimensional gas of $N$ Bosons with point-wise repulsive interaction. Specifically, we consider the protocol \cite{DWBC14} where the system, initially prepared in its non-interacting ground state, is brought out of equilibrium by suddenly turning on a repulsive interaction between the particles.

The aim of this paper is threefold. First, by means of a case study, we show that the quench action method predicts a statistics $p(w)$ of the large deviations of the intensive variable $w$ which takes the exponential form $p(w) \sim {\rm exp}[-L I(w)]$ for $L \rightarrow \infty$ at fixed density $D=N/L$, i.e., $p(w)$ naturally satisfies the so-called \textit{large deviation principle} \cite{Touc09}. In addition, the quench action method allows the calculation of $I(w)$. Second, from this result we carry out a quantitative analysis of the interaction quench described above, pointing out its most interesting features (strongly depending on the presence of the interactions) which cannot be captured by the qualitative picture presented in Ref.~\cite{GaSi12}. Third, going beyond large deviation theory, we analyze the region $w>\bar{w}$ where $I(w)$ vanishes identically and $p(w)$ has an algebraic decay upon increasing $w$. Although, in this case, the quench action method is not sufficient to quantitatively describe $p(w)$, we are able to determine the exponent of this algebraic tail by performing a finite-size calculation in the limit of vanishing densities $D$ of the Bosons. To the best of our knowledge, this provides the first quantitative description of the ``condensed regime'' characterizing the work statistics of interacting Bosonic systems for quenches starting from a critical initial state. Our results are expected to be relevant for experimental realizations of one-dimensional Bose gases in cold-atomic settings \cite{KiWW04,KiWW06,AEWK08,HGMD09,EWAR10,KHML10,FCFF11,DBAD12,FPCF15} and for measurements of the work statistics in non-equilibrium protocols \cite{HSKDL08,DCHFGV13,MDCP13,CBKZH13,Batalhaoetal,Anetal}.

The rest of this paper is organized as follows. In Sec.~\ref{sec:general_setting} we introduce the Lieb-Liniger Hamiltonian and its solution via the Bethe ansatz (Subsecs.~\ref{sec:hamiltonian} and \ref{sec:thermodynamics}), together with details on the quench protocol and the quench action approach (Subsecs.~\ref{sec:quench} and \ref{sec:quench_action}). In Sec.~\ref{sec:work_statistic}, we recall known results on the statistics of the work done in a quantum quench. In Sec.~\ref{sec:exact_rate_functions} we show that the quench action method naturally leads to a probability distribution $p(w)$ of the intensive work which takes the typical form of the large deviation principle, i.e., $p(w) \sim {\rm exp}[-L I(w)]$, where $I(w)$ can be computed within this formalism. Section \ref{sec:general_sec_results} contains all the results regarding the rate function obtained via the quench action approach, while in Sec.~\ref{sec:power_law_tail} we determine the power-law decay characterizing the fluctuations exceeding the mean value $\bar{w}$  in the limit of vanishing densities of the Bosons. Finally, we report our conclusions in Sec.~\ref{sec:conclusions}, while the most technical aspects of our work are presented in several appendices. 

\section{The model and the quench protocol}
\label{sec:general_setting}

\subsection{The Hamiltonian}
\label{sec:hamiltonian}
We consider the Lieb-Liniger model \cite{Lieb63,LiLi63} describing a gas of $N$ Bosons at positions $\{x_1,...,x_N \}$ with mass $m$ and point-wise repulsive interactions, with Hamiltonian
\be
H(c)=-\frac{\hbar^2}{2m}\sum_{j=1}^{N}\frac{\partial^2}{\partial x_{j}^2}+2c\sum_{j<k}\delta(x_j-x_k).
\label{eq:hamiltonian}
\ee
The interaction strength $c$ is related to the scattering length $a_{\rm 1D}$ in one dimension through $c=-\hbar^2/ma_{\rm 1D}$ \cite{olshanii-98} and it can be varied via Feshbach resonances \cite{iasm-98}. In the following we set $\hbar=2m=1$ and assume that the $N$ Bosons are confined within a one-dimensional ring of length $L$, realizing periodic  boundary conditions. 

The Hamiltonian in Eq.~\eqref{eq:hamiltonian} is integrable and, as a consequence, it can be diagonalized exactly by means of the Bethe ansatz \cite{Lieb63}. In particular, the $N$-particle eigenfunctions can be written as
\bea
\psi_N\left(x_1,\ldots,x_N\right)&=& \sum_P \prod_{\ell > k }^{N} \left[1- \frac{i  c
~\text{sgn}(x_\ell - x_k)}{\lambda_{P_\ell} - \lambda_{P_k}}\right]\nonumber\\
&\times & \prod_{j=1}^N e^{i
\lambda_{P_j} x_j}\,,
\label{eq:wave_function}
\eea
where the sum runs over the $N!$ permutations $P$ of $N$ elements. The parameters $\{\lambda_j\}_{j=1}^{N}$ are the so-called rapidities and, in analogy with the quasi-momenta which are relevant in the case of free quantum gases, they parametrize the different eigenstates of the Hamiltonian. When the system has a finite extension $L$, the rapidities have to satisfy a set of quantization conditions which are known as Bethe equations \cite{LiLi63}
\be
e^{-i\lambda_jL}=\prod_{k\neq j}^{N}\frac{\lambda_k-\lambda_j+i{c}}{\lambda_k-\lambda_j-i{c}}\ ,\quad \mbox{with} \quad j=1,\ldots, N\,.
\label{eq:bethe_eq}
\ee
For repulsive interactions $c>0$, it can be shown that all $\lambda_j$'s are real; accordingly, it is convenient to consider the logarithm of Eq.~\eqref{eq:bethe_eq}, i.e.,
\be
\lambda_j=\frac{2\pi I_j}{L}-\frac{2}{L}\sum_{k=1}^N\arctan\left(\frac{\lambda_j-\lambda_k}{c}\right)\, ,
\label{eq:bethe_equations_log}
\ee
where we introduced the quantum numbers $I_j$. These numbers parametrize the sets of rapidities $\{\lambda_j\}$ and are integers (half-integers) for odd (even) $N$; note that they have to be chosen in such a way that $I_j\neq I_k$ for $j\neq k$ \cite{korepin_book}. The knowledge of the rapidities $\{\lambda_j\}_{j=1}^{N}$ completely specifies the eigenstates $\psi_N$ of $H(c)$ and their properties. For example, the corresponding energy eigenvalue can be written as
\be
E\left[\{\lambda_j\}_{j=1}^N\right]=\sum_{j=1}^{N}\lambda_j^2\,.
\label{eq:energy}
\ee
In the following, we will denote the normalized eigenstate of $H(c)$ corresponding to a set of rapidities $\{\lambda_j\}$ as $|\{\lambda_j\}\rangle$.

\subsection{The thermodynamic description}
\label{sec:thermodynamics}

When the number $N$ of particles is very large, the explicit form of the wave function in Eq.~\eqref{eq:wave_function} becomes difficult to deal with, and the Bethe equations \eqref{eq:bethe_equations_log} harder to solve numerically. For these reasons, in order to study the thermodynamic limit of the model, it is necessary to employ an appropriate ``thermodynamic Bethe ansatz'' formalism \cite{YaYa69}, which we briefly review in this section. Here we only report the aspects that are directly relevant to the present study, while the interested reader is referred to Ref.~\cite{takahashi_book} for a thorough treatment.

In the thermodynamic limit $L,N\to\infty$ with fixed density $D=N/L$, it can be seen that the rapidities $\lambda_j$ of a given eigenstate, whose number also grows to infinity, arrange themselves on the real line according to a smooth distribution function $\rho(\lambda)$, with $\lambda\in(-\infty, +\infty)$. Complementary to the latter, one can also introduce a distribution of ``holes'' $\rho^h(\lambda)$, namely of unoccupied states. The functions $\rho(\lambda)$ and $\rho^h(\lambda)$ are analogous to the distributions of momenta and vacancies for free Fermi gases. However, contrary to the non-interacting case, $\rho$ and $\rho^h$ are related in a non-trivial way. In particular, they satisfy the following integral equation
\begin{equation}
\rho^t(\lambda) = \frac{1}{2 \pi} + \frac{1}{2 \pi} \int_{-\infty}^{\infty} {\rm d} \mu \, K(\lambda-\mu) \rho(\mu)
 \label{eq:thermo_bethe}\,,
\end{equation}
where we defined the total distribution function
\be
\rho^t(\lambda)=\rho(\lambda)+\rho^h(\lambda)\,,
\ee
and the kernel 
\begin{equation}
K(\lambda) = \frac{2c}{\lambda^2 + c^2}\,.
 \label{eq:k_function}
\end{equation}  
Equation \eqref{eq:thermo_bethe} can actually be derived by taking the thermodynamic limit of the Bethe equations \eqref{eq:bethe_equations_log} \cite{takahashi_book}. For future use, we also introduce the following standard definition
\bea
\eta(\lambda)=\frac{\rho^h(\lambda)}{\rho(\lambda)}\,. 
\label{eq:eta_function}
\eea
It is widely believed that the knowledge of the rapidity distribution function $\rho(\lambda)$ is sufficient to compute all of the thermodynamic properties of the corresponding eigenstate. For example, in the thermodynamic limit, the densities $D[\rho]$ and $e[\rho]$ of particles and energy per unit length can be obtained, respectively, as
\bea
D[\rho]=\lim_{N,L\to\infty}\frac{N}{L}=\int^{+\infty}_{-\infty}{\rm d}\lambda\,\rho(\lambda)\,,  \label{eq:density_thermo}\\
e[\rho]=\lim_{N,L\to\infty}\frac{E}{L}=\int^{+\infty}_{-\infty}{\rm d}\lambda\,\rho(\lambda)\lambda^2\,.\label{eq:energy_thermo}
\eea

One of the advantages of the thermodynamic description introduced above is the possibility to replace discrete sums over eigenstates with functional integrals over rapidity distribution functions. This is best illustrated by the computation of the thermal partition function at temperature $T= \beta^{-1}$ with $k_B=1$, i.e., 
\be
\mathcal{Z}(\beta)={\rm tr}\left[e^{-\beta H}\right]=\sum_{\{\lambda_j\}}e^{-\beta E\left[\{\lambda_j\}\right]}\,.
\label{eq:thermal_partition}
\ee
Note that, while each term on the right-hand side is known, the sum runs over all the possible sets of rapidities $\{\lambda_j\}$ and hence it is very difficult to evaluate in practice for large $N$. In the thermodynamic limit, however, one can rewrite Eq.~\eqref{eq:thermal_partition} as a functional integral \cite{takahashi_book}
\be
\mathcal{Z}(\beta)=\int \mathcal{D}\rho\,e^{-L \, S_{\rm th}[\beta,\rho]}\,,
\label{eq:thermal_functional_evaluation}
\ee
where the functional
\bea
S_{\rm th}[\beta,\rho]&=&\beta e[\rho]-S_{\rm YY}[\rho]\nonumber\\
&=& \int_{-\infty}^{+\infty}{\rm d}\lambda\,\left\{\beta \rho(\lambda) \lambda^2-s_{YY}[\rho](\lambda)\right\}\,,
\label{eq:thermal_free_energy}
\eea
plays the role of a thermal free energy. Here the first term on the right-hand side is derived using Eq.~\eqref{eq:energy_thermo} for the thermodynamic limit of the energy in Eq.~\eqref{eq:energy} and corresponds to the exponential in Eq.~\eqref{eq:thermal_partition}. The second term, instead, is the so-called Yang-Yang entropy \cite{YaYa69} $S_{\rm YY}=\int{\rm d}\lambda \, s_{YY}(\lambda)$ with density 
\bea
s_{YY}[\rho](\lambda)&=&\rho^t(\lambda)\ln\rho^t(\lambda)-\rho(\lambda)\ln\rho(\lambda)\nonumber\\
&-&\rho^h(\lambda)\ln\rho^h(\lambda)\,, \label{eq:yang_yang_entropy}
\eea
which accounts for the fact that each rapidity distribution function $\rho(\lambda)$ emerges from several ``microscopic realizations'', i.e., that there are many sets of rapidities $\{\lambda_j\}$ associated with the same function $\rho(\lambda)$ \cite{takahashi_book}. For $L \rightarrow \infty$ the functional integral in Eq.~\eqref{eq:thermal_functional_evaluation} can be computed by a saddle-point evaluation, yielding the following expression for the free energy density $f$ associated with the thermal partition function in Eq.~\eqref{eq:thermal_partition} \cite{takahashi_book,korepin_book}
\be
f= -\frac{ T \, \rm{ln \, \mathcal{Z}}}{L} = D h -\frac{T}{2 \pi} \int_{-\infty}^{\infty} {\rm d} \lambda \, \rm{ln}\left(1+\rm{e}^{-\varepsilon(\lambda)/T}  \right)\,. \label{eq:thermal_free_energy_2}
\ee
Here $\varepsilon(\lambda)$ is the solution to the integral equation  
\bea
\varepsilon(\lambda) &=& \lambda^2 -h\nonumber\\
& -&\frac{T}{2 \pi} \int_{-\infty}^{+\infty} {\rm d} \mu \, K(\lambda -\mu) \rm{ln}\left(1+e^{-\varepsilon(\lambda)/T}   \right)\,, 
\label{eq:thermal_dressed_energy}
\eea
and $h$ is a Lagrange multiplier introduced in order to enforce the assigned density of particles, according to Eq.~\eqref{eq:density_thermo}. As we will see in Sec.~\ref{sec:quench_action}, the formalism discussed above will also be essential for introducing the quench action approach.

\subsection{The quench protocol}
\label{sec:quench}
As anticipated, in this work we consider an interaction quench in which the system is initially prepared in the ground state of the non-interacting Hamiltonian $H(c_0=0)$, usually denoted by $|\rm{BEC} \rangle$. The corresponding wave-function $\psi_N^{(0)}(x_1,x_2\ldots,x_N)=\langle x_1,x_2,\ldots\,,x_N| \rm{BEC} \rangle$  reads
\begin{equation}
\psi_N^{(0)}(x_1,x_2...x_N) = \frac{1}{L^{N/2}}\,.
\label{eq:BEC_wave_function}
\end{equation}                          
At time $t=0$, a finite inter-particle repulsive interaction $c>0$ is turned on and the gas is subsequently left to evolve unitarily. Our motivation to investigate the above quench is twofold: first, the simplicity of the initial state allows one to derive analytic predictions which would be difficult to obtain in general; second, we will see that this kind of quench leads to interesting features in the work statistics.

The non-equilibrium dynamics arising from the interaction quench $c_0 \rightarrow c$ described above has been extensively investigated in the literature \cite{GrRD10,KSCC13,KoCC14,DeCa14,DePC15,PiCE16,ZWKG15}. From the analytical point of view, an important result has been the discovery in Ref.~\cite{DWBC14} of an exact formula (later proven in Ref.~\cite{Broc14}) for the overlaps between the initial state in Eq.~\eqref{eq:BEC_wave_function} and the eigenstates of the Hamiltonian in Eq.~\eqref{eq:wave_function}. This formula, which is an essential ingredient for the application of the quench action approach described in Sec.~\ref{sec:quench_action}, will be used several times in this work, and is hence reviewed in what follows.

It was first shown in Refs.~\cite{DWBC14,Broc14} that the initial state in Eq.~\eqref{eq:BEC_wave_function} has a non-vanishing overlap only with eigenstates corresponding to sets $\{\lambda_j\}$ of rapidities which are parity invariant, i.e., such that $\{\lambda_j\}=\{-\lambda_j\}$. This implies that the set $\{\lambda_j\}$ can be decomposed as
\bea
\{\lambda_j\}_{j=1}^{N}=\{\lambda^{+}_j\}_{j=1}^{N/2}\cup \{-\lambda^{+}_j\}_{j=1}^{N/2}\,,
\eea
if $N$ is even and 
\bea
\{\lambda_j\}_{j=1}^{N}=\{\lambda^{+}_j\}_{j=1}^{(N-1)/2}\cup \{-\lambda^{+}_j\}_{j=1}^{(N-1)/2}\cup \{0\} \,,
\eea
if $N$ is odd, where $\lambda^{+}_j>0$. For these states, the overlap formula is extremely simple. Explicitly, for even $N$, it reads
\be
\langle  \{\lambda_j\} | {\rm BEC} \rangle = \sqrt{ \frac{(cL)^{-N}N!}{ \det_{j,k=1}^{N}G_{jk}  } }  \frac{\det_{j,k=1}^{N/2} G^{Q}_{jk}}{{\displaystyle \prod\limits_{j=1}^{N/2} \frac{\lambda_j}{c}   \sqrt{\frac{\lambda_j^2}{c^2} + \frac{1}{4} } }}\,.
\label{eq:overlaps}
\ee
Here we introduced the matrices $G_{jk}$ and $G_{jk}^{Q}$, with elements
\bea
G_{jk} &=& \delta_{jk} \Big[ L + \sum_{l=1}^{N/2}\, K(\lambda_{j}-\lambda_{l}) \Big] - K (\lambda_{j}-\lambda_{k})\,,
\label{eq:Gaudin}\\
G^Q_{jk} &=& \delta_{jk} \Big[ L + \sum_{l=1}^{N/2}\, K^Q(\lambda_{j},\lambda_{l}) \Big] - K^Q (\lambda_{j},\lambda_{k})\,,
\label{eq:Gaudin_Q}
\eea
where $K^{Q}(\lambda,\mu) = K(\lambda - \mu)+ K(\lambda + \mu)$, and $K(\lambda) $ is defined in Eq.~\eqref{eq:k_function}. An analogous result holds for the case of odd $N$ \cite{BDWC14}. As we will see further below, our analytic study ultimately hinges on the existence of the exact formula~\eqref{eq:overlaps}.

\subsection{The quench action method}
\label{sec:quench_action}

In this section we discuss the quench action approach, which is the last piece of technical background needed in order to carry out our analysis of the statistics of the work. In the following, we only review some relevant aspects, referring the reader to the literature for a more comprehensive treatment \cite{Caux16}.

This integrability-based method has been introduced in Ref.~\cite{CaEs13} to tackle the difficult problem of computing the thermodynamic limit of time averages 
\be
\langle\psi_0|\mathcal{O}(t)|\psi_0\rangle=\sum_{n,m} \langle n|\mathcal{O}|m\rangle \langle\psi_0|n\rangle \langle m|\psi_0\rangle  e^{i(E_n-E_m)t}\,,
\label{eq:time_average}
\ee
after a global quench, where $|\psi_0\rangle$ is the initial state and $\mathcal{O}$ a generic local observable. Here we denoted by $|n\rangle$ and $|m \rangle$ the eigenstates of the Hamiltonian with energies $E_n$ and $E_m$, respectively. While, in principle, the quench action approach can be used to compute the full evolution of the expectation value in Eq.~\eqref{eq:time_average} \cite{DeCa14,DePC15,BeSE14,PiCa17}, it is particularly effective if one is interested only in its infinite-time limit~\cite{Caux16}
\bea
\lim_{t\to\infty}\langle\psi_0|\mathcal{O}(t)|\psi_0\rangle=\sum_n |\langle n|\psi_0\rangle|^2 \langle n|\mathcal{O}|n\rangle \nonumber\\
=\sum_{\{\lambda_j\}} |\langle {\{\lambda_j\}}|\psi_0\rangle|^2 \langle {\{\lambda_j\}}|\mathcal{O}|{\{\lambda_j\}}\rangle\,,
\label{eq:infinite_limit}
\eea
where we used the fact that for the Lieb-Liniger model the eigenstates are parametrized by sets of rapidities $\{\lambda_j\}$. Indeed, the quench action approach provides a simple prescription to evaluate the thermodynamic limit of Eq.~\eqref{eq:infinite_limit}, which is based on replacing the spectral sum
\be
\sum_{\{\lambda_j\}} |\langle {\{\lambda_j\}}|\psi_0\rangle|^2
\label{eq:discrete_overlap_sum}
\ee
in Eq.~\eqref{eq:infinite_limit} with a functional integration, in analogy with what we did in Eq.~\eqref{eq:thermal_functional_evaluation}. By doing so, one arrives at the formal expression
\bea
\lim_{t\to\infty}\langle\psi_0|\mathcal{O}(t)|\psi_0\rangle= \int \mathcal{D}\rho\,\langle \rho |\mathcal{O}|\rho\rangle e^{-L \, S_{\rm \rm QA}[\rho]}\,,
\label{eq:QA_final_result}
\eea
where we denoted by $|\rho\rangle$ an eigenstate whose rapidities $\{\lambda_j\}$ approach the distribution $\rho(\lambda)$ in the thermodynamic limit. Here, the functional $S_{\rm QA}[\rho]$, usually called the quench action (QA), plays a role analogous to the thermal free energy in Eq.~\eqref{eq:thermal_free_energy}. Explicitly, it reads \cite{Caux16}
\bea
S_{\rm QA}[\rho] &=& 2 S_O[\rho] -\frac{1}{2}S_{\rm YY}[\rho] + S_N[\rho] \nonumber\\
&=&2S_{O}[\rho] 
-  \frac{1}{2}\int_{-\infty}^{+\infty}{\rm d}\lambda\,s_{YY}[\rho](\lambda)\nonumber\\
 &+&\frac{h}{2}\left[\int_{-\infty}^{+\infty}{\rm d}\lambda\,\rho(\lambda)-D\right]\,, 
\label{eq:quench_action} 
\eea
where $S_{O}[\rho]$ is the functional associated with the overlap term in the spectral sum in Eq.~\eqref{eq:discrete_overlap_sum}, i.e.,
\be
|\langle {\{\lambda_j\}}|\psi_0\rangle| \simeq e^{-L \, S_{O}[\rho]}\,.
\label{eq:overlap_functional}
\ee
Note that, since the Hamiltonian in Eq.~\eqref{eq:hamiltonian} conserves the particle number which is well-defined in the initial state, a Lagrange multiplier  $h$ has been introduced in Eq.~\eqref{eq:quench_action} (where the prefactor $1/2$ is for later convenience): this allows us to extend the functional integration over the whole space of rapidity distribution functions. Note also that the Yang-Yang entropy appearing in Eq.~\eqref{eq:quench_action} bears an additional prefactor $1/2$, which is due to the fact that only parity-invariant eigenstates contribute to the spectral sum in Eq.~\eqref{eq:discrete_overlap_sum} \cite{DWBC14}. The integral in Eq.~\eqref{eq:QA_final_result} can now be computed via the saddle-point method, yielding an exact result for the infinite-time average of the expectation value of local observables in the thermodynamic limit.

A crucial point in the procedure outlined above is the availability of an analytic expression for the functional $S_O[\rho]$ in Eq.~\eqref{eq:overlap_functional}. If an analytic expression of the form \eqref{eq:overlaps} is known for the overlaps, that for $S_O[\rho]$ can be easily derived as shown explicitly in Ref.~\cite{DWBC14}; in particular, for the case of the initial state in Eq.~\eqref{eq:BEC_wave_function} one has
\bea
S_{O}[\rho]&=&\frac{D}{2}\left[1+\ln\gamma\right]\nonumber\\
&+&\frac{1}{4}\int_{-\infty}^{+\infty}{\rm d}\lambda \, \rho(\lambda)\ln\left[\frac{\lambda^2}{c^2}\left(\frac{\lambda^2}{c^2}+\frac{1}{4}\right)\right]\,,
\label{eq:BEC_overlap_term}
\eea
where we introduced the normalized interaction strength
\be
\gamma=\frac{c}{D}\,.
\ee
Unfortunately, for arbitrary initial states it remains an open problem whether formulas analogous to Eq.~\eqref{eq:overlaps} can be derived, so that, in general, the explicit expression for $S_O[\rho]$ is unknown \cite{KoPo12,Pozs14,PiCa14,BNWC14,LeKZ15,LeKM16,Pozs18,LeKL18}. Note, however, that it was recently shown that for any integrable model it is always possible to find a class of ``integrable initial states'' for which this can be done \cite{PiPV17,PoPV18}. As we will comment on later, the results derived in this work can thus be generalized straightforwardly to other integrable systems, at least for quenches from the latter class of initial states.

\section{The statistics of the work}
\label{sec:work_statistic}

Before presenting our results, we review some generic features of the work statistics obtained in free models \cite{TalLutz07,Silv08,GaSi11,GaSi12,SoGS13,SmSi13}. The following discussion will be useful for a comparison with the interacting case analyzed in this paper.

Consider a quantum system with $N$ degrees of freedom, initially in the ground state $| \psi_0 \rangle$ of its Hamiltonian $H(c_0)$. In the following, we denote its ground-state energy eigenvalue with $E_0^{c_0}$. The probability distribution $P(W)$ of the extensive work $W$ done in quenching a global parameter $c_0 \rightarrow c$ is defined as \cite{TalLutz07}  
\be 
P(W) = \sum_{n \geq 0} |\langle \psi_n^{c}| \psi_0 \rangle|^2 \delta(W-(E_n^{c}-E_0^{c_0})), \label{eq:work_probability_definition}
\ee
where $| \psi_n^{c} \rangle$ are the eigenstates of the post-quench Hamiltonian $H(c)$ with corresponding energies $E_n^{c}$. One immediately notices from Eq.~\eqref{eq:work_probability_definition} that the work has a minimum threshold value $W_{\rm{rev}}=E_0^{c}-E_0^{c_0}$. This has the meaning of reversible work, i.e., the work performed at zero temperature when the transformation $c_0 \rightarrow c$ is done in a reversible way. As a consequence, we will refer the work to this threshold, focussing on the irreversible contribution $W_{\rm{irr}}= W-W_{\rm{rev}} \geq 0$, which is related to the irreversible entropy production \cite{PolkSeng11,DorGoold12}. For convenience, we will henceforth indicate $W_{\rm{irr}}$ by $W$, dropping the subscript.      

In the following, we will be interested in the moment generating function $G(s)$ of the extensive work $W>0$
\be
G(s) = \langle \mbox{e}^{-s W} \rangle = \langle \psi_0| \mbox{e}^{-s(H(c)-E_0^{c})} | \psi_0 \rangle , \label{eq:G(s)_definition}   
\ee
and in the corresponding scaled cumulant generating function (SCGF) $f(s)$, defined by  \cite{Touc09} 
\be
G(s) = \mbox{e}^{-N f(s)}\,. 
\label{eq:ldev_f(s)}
\ee
Note that, from Eq.~\eqref{eq:G(s)_definition}, $\langle W \rangle = \langle \psi_0| H(c) |\psi_0 \rangle -E_0^c \,,$ which motivates the expectation that $\langle W \rangle \propto N$. As anticipated in the introduction, in order to investigate the large deviations of the random variable $W$, it is convenient to focus on the intensive irreversible work $w=W/N$, with probability density $p(w)$. Upon increasing $N$ one generically expects that $p(w)$ satisfies the so-called large deviation principle \cite{Touc09}, i.e., that $p(w) \sim {\rm exp}[-N I(w)]$, where $I(w)$ is referred to as the rate function. 
This function is non-negative, convex and, in general, displays a unique zero at the average and most probable value $ \bar{w}$: upon increasing $N$ the function $p(w)$ becomes peaked around $\bar{w}$, being exponentially suppressed for $w\neq \bar{w}$. Importantly, $I(w)$ can be computed by means of the G\"{a}rtner-Ellis theorem, which states that $I(w)$ is given by the Legendre-Fenchel transform of $f(s)$ in Eq.~\eqref{eq:ldev_f(s)}, namely
\be 
I(w) = -\mbox{inf}_{s}\{ s w -f(s)  \}\,, 
\label{eq:proved_Gartner}
\ee     
where the infimum has to be taken within the domain in which $f(s)$ is defined. Note that, once the large deviation principle $p(w) \sim {\rm exp}[-N I(w)]$ is satisfied, the G\"{a}rtner-Ellis theorem can be heuristically derived by a saddle-point approximation of the inverse Laplace transform of $G(s)$ \cite{GaSi12,Touc09}. However, it might be difficult to prove this principle a priori in specific cases, and thus this is usually done a posteriori.

In the case of free (Fermionic and Bosonic) models, the rate function is quadratic in a neighborhood of $\bar{w}$, so that small deviations from the average intensive work have a Gaussian distribution \cite{GaSi12}: this is what one would expect from a naive application of the central limit theorem. On the contrary, for large deviations from $\bar{w}$, $I(w)$ might differ significantly from its quadratic approximation, displaying interesting features. Most prominently, as shown in Refs.~\cite{GaSi12,SoGS13}, the behavior of $I(w)$ for $w \ll \bar{w}$ becomes universal if the post-quench Hamiltonian is close to criticality, a fact which can be rationalized via a quantum-to-classical correspondence. 

In the case of systems composed by free Bosonic excitations, it has been shown in Ref.~\cite{GaSi12} that $p(w)$ may feature a different kind of universal behavior also for $w>\bar{w}$. In particular, as the pre-quench initial state is varied from being non-critical to critical, a transition in the form of $p(w)$ takes place such that $I(w)$ vanishes identically for $w \geq \bar{w}$ when the pre-quench initial state becomes critical. This has been identified as a ``condensation'' transition, in analogy to the Bose-Einstein condensation in equilibrium statistical mechanics \cite{Huang} and it implies that $p(w)$ displays an algebraic decay upon increasing $w$. Although no general expression has been reported so far for such a power-law tail of $p(w)$ in this condensation regime, the latter has been shown to appear also in different non-equilibrium protocols \cite{SmSi13}. In this work we will provide a quantitative prediction for the corresponding exponent in the interaction quench introduced in Sec.~\ref{sec:quench}.

\section{From the quench action approach to large deviation theory: the rate function}
\label{sec:exact_rate_functions}

We now present our analysis and predictions for the statistics of the work done by the quench introduced in Sec.~\ref{sec:quench}. We begin by showing that the quench action approach allows us to demonstrate, directly and rather generally, that the large deviation principle $p(w)\sim e^{-LI(w)}$ holds and then compute the rate function $I(w)$.

We start from the expression of the moment generating function $G(s)$ in Eq.~\eqref{eq:G(s)_definition} in which we insert the resolution of the identity operator $\mathbb{I}$ in terms of the post-quench Bethe eigenstates $|\{\lambda_j\}\rangle$
\begin{equation}
\mathbb{I} = \sum_{\{\lambda_j \}} |\{\lambda_j\}\rangle \langle \{\lambda_j\} |,
\end{equation}
obtaining
\begin{equation}
G(s) =  \sum_{\{\lambda_j \}} |\langle \{\lambda_j\} |\mbox{BEC} \rangle|^2 \mbox{e}^{-s(E[\{\lambda_j\}]-E_0^c)}. \label{eq:intermediate_G(s)_QA}
\end{equation}
One then notices that Eq.~\eqref{eq:intermediate_G(s)_QA} has a structure analogous to that of Eq.~\eqref{eq:infinite_limit}. Accordingly, it can be expressed as the r.h.s of Eq.~\eqref{eq:QA_final_result} which involves the quench action $S_{\rm QA}[\rho]$, namely
\begin{eqnarray} 
G(s) & = & \int \mathcal{D \rho} \, \, \mbox{exp}[-L S_{\rm QA}[\rho] -s(E[\{\lambda_j\}]-E_0^c)] \nonumber \\
     & = & \int \mathcal{D \rho} \, \, \mbox{exp}[-L (S_{\rm QA}[s,\rho]-s e_0(c))] \,, \label{eq:QA_saddle_G(s)} 
\end{eqnarray}
where we introduced the ground-state energy density $e_{0}(c) = E_{0}^c/L$, and the modified quench action
\be
S_{\rm QA}[s,\rho] = S_{\rm QA}[\rho] +s \, e[\rho],  \label{eq:modified_quench_action}
\ee
with $S_{\rm QA}[\rho]$ given in Eq.~\eqref{eq:quench_action}, and $e[\rho]$ in Eq.~\eqref{eq:energy_thermo}. In the thermodynamic limit, the functional integral in Eq.~\eqref{eq:QA_saddle_G(s)} can be evaluated via the saddle-point method, leading to 
\begin{equation}
G(s)  \sim  \mbox{exp}[-L(S_{\rm QA}[s,\rho_{s}^{\ast}]-s e_0(c))]\,.
\label{eq:functional_integral_quench_action_2}
\end{equation}
Here the function $\rho^{\ast}_{s}$ is determined by the saddle-point condition
\begin{equation}
\frac{\delta S_{\rm QA}[s,\rho]}{\delta \rho}\Bigr|_{\rho=\rho_s^{\ast}} \equiv 0\,.
\label{eq:saddle_point}
\end{equation}
Note that by straightforward manipulations, Eq.~\eqref{eq:saddle_point} can be cast into the explicit form [see also Eqs.~\eqref{eq:quench_action}, \eqref{eq:BEC_overlap_term}, and \eqref{eq:yang_yang_entropy}]
\begin{equation}
\begin{split}
\varepsilon_s^{\ast}(\lambda) &=  2  \lambda^2  + \frac{1}{s}\mbox{ln}\left[ \frac{\lambda^2}{c^2}\left( \frac{\lambda^2}{c^2} +\frac{1}{4}    \right)\right] \\
&-\frac{h}{s}- \frac{1}{s}\int_{-\infty}^{\infty}\frac{{\rm d} \mu}{2 \pi} K(\lambda-\mu) \, \mbox{ln}\left(1+ \mbox{e}^{-s \varepsilon_s^{\ast}(\mu)} \right),
\label{eq:saddlepoint_density}
\end{split}
\end{equation}
involving, instead of $\rho_s^{\ast}$,
\be
\varepsilon_s^{\ast}(\lambda)= \frac{1}{s}\mbox{ln} \, \eta_s^{\ast}(\lambda)\,,
\ee
where $\eta(\lambda)$ is defined in Eq.~\eqref{eq:eta_function}. Equation \eqref{eq:saddlepoint_density} has to be interpreted as follows. For each value of $s$, one finds a unique solution for the function $\varepsilon_s^\ast(\lambda)$, and hence for $ \eta_s^{\ast}(\lambda)$. Then, by recalling that $\rho^{t}(\lambda)=\rho(\lambda)(1+\eta(\lambda))$, one plugs the latter function into Eq.~\eqref{eq:thermo_bethe}, in order to obtain a final prediction for $\rho^{\ast}_s(\lambda)$. Note that the Lagrange multiplier $h(s)$ in Eq.~\eqref{eq:saddlepoint_density} has to be chosen such that the prescribed density $D$ is obtained after using Eq.~\eqref{eq:density_thermo}. 

Within the saddle-point approximation in Eq.~\eqref{eq:functional_integral_quench_action_2} one finds that, from Eq.~\eqref{eq:ldev_f(s)}, 
\be
f(s) = -\frac{1}{L}\mbox{ln} \, G(s) = S_{\rm QA}[s,\rho_s^{\ast}]-s e_0(c) \label{eq:proved_SCGF}
\ee 
and therefore, in order to calculate $I(w)$ according to Eq.~\eqref{eq:proved_Gartner} one has to find the infimum, as a function of $s$, of $s w -f(s)$. When this is attained in a differentiable point $s_w$, it is determined by the condition 
\be
\frac{\rm d}{{\rm d}s}\left(f(s)-sw\right)\Big|_{s=s_w}=0\,.
\label{eq:derivative_condition}
\ee
Due to the concavity of $f(s)$ \cite{Touc09}, the stationary point $s=s_w$ can only correspond to a minimum. Using now $({\rm d}/{\rm d}s)=({\rm d\rho}/{\rm d}s)(\delta/\delta\rho)$, and exploiting Eq.~\eqref{eq:saddle_point}, one can easily show that this condition is in fact equivalent to requiring
\begin{equation}
\int_{-\infty}^{+\infty}{\rm d\lambda} \, \rho^\ast_{s_w}(\lambda) \lambda^2-e_{0}(c) = w\,. 
\label{eq:energyconstraint}
\end{equation} 
As a consequence, if $f(s)$ is in addition strictly concave, the expression in Eq.~\eqref{eq:proved_Gartner} simplifies as 
\be
I(w)=-s_w w+f(s_w)\,. 
\label{eq:Lengendre_duality}
\ee
Importantly, in this derivation, we never had to evaluate the quench action $S_{\rm QA}[s,\rho]$ at complex values of $s$, where it has been shown that it might display singular points \cite{PiPV17_II}. Note also that the specific form of the overlap term $S_O[\rho]$, entering only in Eq.~\eqref{eq:saddlepoint_density}, does not play any role in this derivation. As a consequence, the latter can be generalized straightforwardly to any integrable model where the quench action approach can be applied. We note that the function $f(s)$ in Eq.~\eqref{eq:proved_SCGF} has been defined with a rescaling by the system size $L$ and not by the number of Bosons $N$ as in Eq.~\eqref{eq:ldev_f(s)}. The two definitions are clearly equivalent since $N=D L$ and the density $D$ is assumed to be fixed.   

We can now proceed towards the explicit evaluation of the rate function $I(w)$, using Eq.~\eqref{eq:proved_Gartner}. First, note that exploiting Eqs.~\eqref{eq:thermo_bethe} and \eqref{eq:saddlepoint_density}, the action in Eq.~\eqref{eq:modified_quench_action} can be rewritten in the compact form
\bea
S_{QA}[s,\rho_s^{\ast}] &=& D \left( \mbox{ln} \gamma  +1 \right) \nonumber\\
&+& \frac{h D}{2} -\frac{1}{2} \int_{-\infty}^{\infty} \frac{{\rm d} \lambda}{2 \pi} \, \mbox{ln} \left( 1+ \mbox{e}^{-s \varepsilon_s^{\ast}(\lambda)} \right).
\label{eq:final_SCGF}
\eea
Next, one needs to solve Eqs.~\eqref{eq:saddlepoint_density} and \eqref{eq:thermo_bethe} with the constraint in Eq.~\eqref{eq:density_thermo}. This can be easily done numerically by standard iterative procedures. The resulting solution for $\varepsilon_s^{\ast}(\lambda)$ can then be plugged into Eq.~\eqref{eq:final_SCGF} and integrated numerically. Finally, in order to obtain $f(s)$ in Eq.~\eqref{eq:proved_SCGF}, one also needs to compute the ground-state energy $e_{0}(c)$. In fact, this can be written in terms of the solution of an integral equation (see, e.g., Ref.~\cite{takahashi_book}). In particular, we have
\begin{equation}
e_{0}(c) = \int_{-Q}^{Q}{\rm d}\lambda \, \rho_{\rm GS}(\lambda)\lambda^2\,,
\end{equation}
where $\rho_{\rm GS}(\lambda)$ satisfies the Lieb equation
\begin{equation}
\rho_{\rm GS}(\lambda)=\frac{1}{2\pi}+\frac{1}{2\pi}\int_{-Q}^{Q} {\rm d \mu} \, K(\lambda-\mu)\rho_{\rm GS}(\mu)\,, \quad |\lambda|<Q\,,
\label{eq:lieb_equation}
\end{equation}
and where the real number $Q$ is determined self-consistently by requiring
\begin{equation}
\int_{-Q}^{Q}{\rm d}\lambda \, \rho_{\rm GS}(\lambda)=D\,.
\label{eq:lieb-constraint}
\end{equation}
We have now all the necessary ingredients to evaluate the rate function $I(w)$, which is obtained by numerically performing the Legendre-Fenchel transform in Eq.~\eqref{eq:proved_Gartner}. The latter expression is indeed better suited for a numerical evaluation of $I(w)$ than Eq.~\eqref{eq:Lengendre_duality} since $f(s)$ is in general known only numerically from Eq.~\eqref{eq:proved_SCGF}. The Legendre transform in Eq.~\eqref{eq:Lengendre_duality} will be instead used in order to determine analytically the asymptotic behavior of $I(w)$ both close to $\bar{w}$ and for low values of $w$. We have implemented the numerical procedure outlined above, which presents no difficulty, and we have worked out analytically the asymptotic of $I(w)$; our results are summarized and discussed in the next section. 

\begin{figure*}
	\centering
	\includegraphics[scale=0.87]{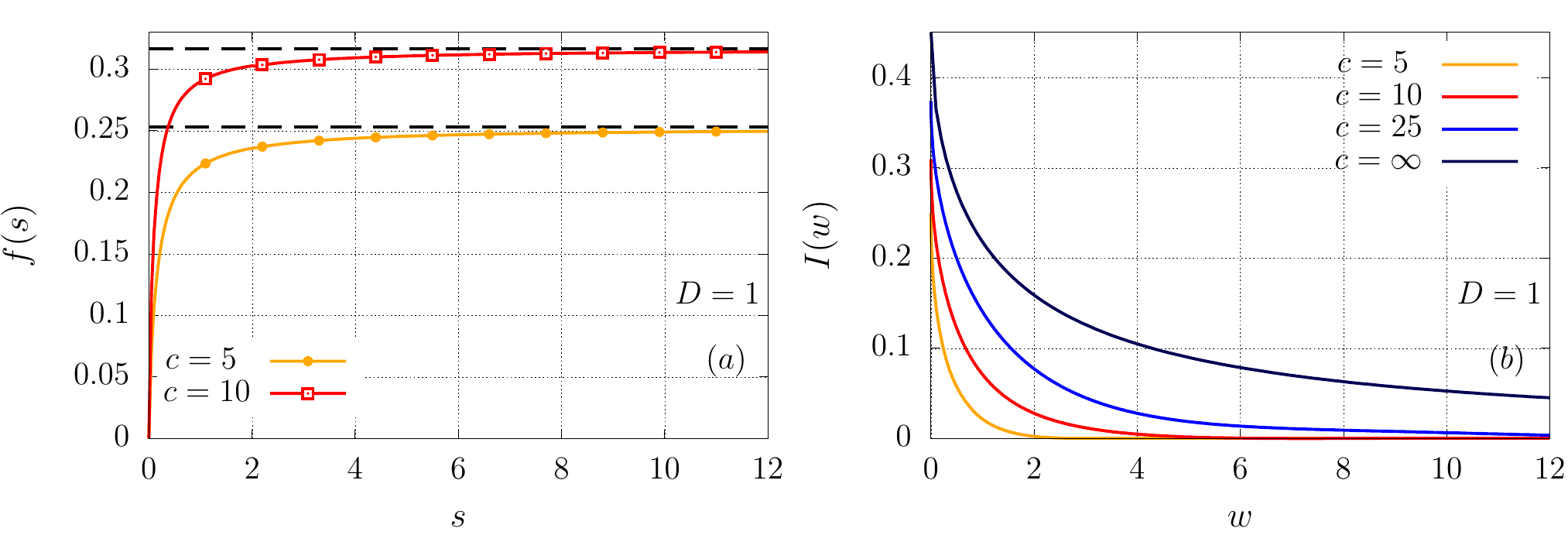}
	\caption{$(a)$ Numerical determination of the function $f(s)$ for various values of the interaction $c=5,10$ (from bottom to top) and fixed density $D=1$. Solid lines correspond to the exact numerical values obtained by solving Eq.~\eqref{eq:saddlepoint_density}, while dashed lines indicate the asymptotic value $2 f_0$ determined from, c.f., Eq.~\eqref{finallargesasymptotic} in Sec.~\ref{sec:asymptotics_calculations}. $(b)$ Large deviation function $I(w)$ for various values of the post-quench interaction $c$. In black we show, for comparison, $I(w)$ in the Tonks-Girardeau limit $c \rightarrow \infty$, evaluated in Sec.~\ref{sec:c_infinity_limit}. The latter is never identically zero, in contrast to the curves corresponding to $c=5,10,25$ (from bottom to top) which vanish identically for $w\geq \bar{w}= c D^2 -e_0(c)$.}
	\label{fig:compare_interactions_f_I}
\end{figure*}

\section{Results}
\label{sec:general_sec_results}
In this section we present our results for the scaled cumulant generating function in Eq.~\eqref{eq:proved_SCGF} and for the rate function $I(w)$. We begin by reporting in Sec.~\ref{sec:numerical_results} their numerical evaluation based on the exact formulas presented in the previous section, and then work out analytically their asymptotic behavior in Sec.~\ref{sec:asymptotics_calculations}. Finally, we devote Sec.~\ref{sec:c_infinity_limit}  to a detailed analysis of the so-called Tonks-Girardeau (TG) limit $c\rightarrow \infty$ , which lends itself to a fully analytical treatment.

\subsection{The exact rate function: numerical results }
\label{sec:numerical_results}

We start by presenting our numerical predictions for the scaled cumulant generating function $f(s)$, which are reported in Fig.~\ref{fig:compare_interactions_f_I}($a$). The data are obtained using Eq.~\eqref{eq:proved_SCGF}, after numerical solution of Eqs.~\eqref{eq:saddlepoint_density}, \eqref{eq:thermo_bethe}, and \eqref{eq:lieb_equation}, which is done by standard iterative procedures \cite{Klauser2011}. 

We see from Fig.~\ref{fig:compare_interactions_f_I}($a$), that $f(s)$ displays many of the generic features predicted in Ref.~\cite{GaSi12}. In particular, it is a concave function defined in a semi-infinite interval $[\bar{s},\infty)$ of the real line. In this case $\bar{s}=0$, since the function $f(s)$ diverges for $s<0$. This is also consistent with the fact that $f(s)$ has a singular point in its second derivative at $s=0$. To see this, one can compute the second derivative of $f(s)$, yielding
\begin{equation}
\frac{{\rm d}^2}{{\rm d}s^2} f(s)\Big|_{s=0} = \langle {\rm BEC}|\left[H(c)-E_{0}(c) \right]^2| {\rm BEC} \rangle\to \infty\,,  \label{infinitevariance}
\end{equation}
as it can be verified by using the Wick theorem and by noticing that divergent terms arise on the r.h.s. of Eq.~\eqref{infinitevariance}.     
On the contrary, the first derivative of $f(s)$ is finite in $s=0$ and it gives the average intensive work $\bar{w}$ performed in the quench. In particular, from Eq.~\eqref{eq:proved_SCGF}, we have
\begin{eqnarray}
\bar{w}= -\frac{G'(0)}{L}=f'(0)  &=&  \frac{1}{L} \, \langle {\rm BEC}| H(c)-E_0(c)| {\rm BEC} \rangle \nonumber \\
                                 &=&  c D^2 -e_{0}(c) \,.
\label{eq:firt_derivative}
\end{eqnarray}
As $s$ approaches zero, $f(s)$ displays a non-analytic behavior, which will be characterized in Sec.~\ref{sec:asymptotics_calculations} and which is responsible for the divergence of the higher-order cumulants. We also note that the qualitative features of $f(s)$ do not change upon varying $c$. However, the average work $\bar{w}$, and hence the first derivative in $s=0$, increases upon increasing the interaction $c$. This is expected because as the repulsion among the Bosons increases, a larger energy is present in the initial BEC state where all the Bosons have zero momentum according to Eq.~\eqref{eq:BEC_wave_function}. In particular, in the Tonks-Girardeau limit $c \rightarrow \infty$ the mean intensive work $\bar{w}$ diverges. 

From Fig.~\ref{fig:compare_interactions_f_I}($a$) one also sees $f(s)$ slowly approaches its asymptotic value for $s\to \infty$, which increases upon increasing $c$. This asymptotic behavior, however, is difficult to analyze numerically and we postpone its discussion to Sec.~\ref{sec:asymptotics_calculations}, where it will be determined analytically.

In Fig.~\ref{fig:compare_interactions_f_I}($b$) we report our predictions for the rate function $I(w)$ corresponding to $f(s)$ in panel ($a$), which can be obtained after numerical Legendre-Fenchel transform of $f(s)$, as explained in the previous section. As we discussed above, the rate function $I(w)$ vanishes at $w=\bar{w}$, while it is identically zero for $w>\bar{w}$. This means that fluctuations for $w>\bar{w}$ must have a sub-exponential dependence $C(L,c)$ on the system size $L$. In fact, assuming a power-law decay 
\begin{equation}
p(w) \sim  C(L,c) \, w^{-\beta}  \: \: \, \mbox{for} \quad w \gg \bar{w}\,, 
\label{powerlawtail}
\end{equation} 
one can constrain the value of $\beta$ by taking into account the divergence of cumulants beyond the first one, see Eqs.~\eqref{infinitevariance} and \eqref{eq:firt_derivative}. In particular, it must be $ 2< \beta <3$. While the pre-factor $C(L,c)$ and the exponent $\beta$ can not be determined from large deviation theory, we will show that they can be calculated from the Bethe ansatz in Sec.~\ref{sec:power_law_tail}, at least in the limit of vanishing densities of Bosons. 

Finally, from Fig.~\ref{fig:compare_interactions_f_I}($b$) we see that $I(w)$ rapidly approaches zero as $w\to \bar{w}^{-}$. In fact, due to the limitations in the accuracy of the numerical solutions, it is difficult to characterize this decay numerically, as $I(w)$ becomes very small when $w\simeq \bar{w}$. However, as we show in the next section, this regime can be successfully tackled analytically, so that the behavior of $I(w)$ near $\bar{w}$ can be determined exactly.

\subsection{Asymptotic behavior of the rate function: analytic results}
\label{sec:asymptotics_calculations}

As we mentioned in Sec.~\ref{sec:work_statistic}, whenever the central limit theorem applies \cite{Touc09} because $w$ can be seen as the sum of a large number of microscopic works done separately on the single particles, the rate function $I(w)$ has a quadratic expansion around $w=\bar{w}$. This is, for instance, the case for free fermionic models \cite{Silv08,GaSi11,GaSi12,SoGS13,SmSi13}, where a Gaussian distribution describes $p(w)$ for small deviations from its mean value $\bar{w}$. In the case under study, however, we show in the following that the behavior of $I(w)$ near $\bar{w}$ is not Gaussian, meaning that the central limit theorem does not apply. 

In order to study the behavior of $I(w)$ for $w \rightarrow \bar{w}^{-}$, we exploit Eq.~\eqref{eq:Lengendre_duality} which applies to our case since $f(s)$ is strictly concave and therefore $f'(s)$ is invertible. Due to the concavity of $f(s)$, it is easy to show that the behavior of $I(w)$ near $\bar{w}$ is determined by the expansion of $f(s)$ in a neighborhood of $s=0$. In other words, we are left with the problem of determining the form of $f(s)$ for small $s$.  To this end, we start from Eq.~\eqref{eq:proved_SCGF} and define
\be
a_s(\lambda)=\frac{1}{\eta_s^{\ast}(\lambda)}\,.
\ee
Differentiating Eq.~\eqref{eq:proved_SCGF} with respect to $s$, we obtain
\be
\frac{{\rm d}}{{\rm d} s}f(s)=\frac{h^\prime(s)D}{2}-\frac{1}{2}\int^{+\infty}_{-\infty}\frac{{\rm d}\lambda}{2\pi}\frac{a^{\prime}_s(\lambda)}{1+a_s(\lambda)}-e_{0}(c)\,.
\ee
Next, differentiating Eq.~\eqref{eq:saddlepoint_density} with respect to $s$, multiplying each side of the resulting equation by $\rho_s^{\ast}(\lambda)$ and finally integrating in $\lambda$ we obtain
\be
\frac{h^\prime(s)D}{2}-\frac{1}{2}\int^{+\infty}_{-\infty}\frac{{\rm d}\lambda}{2\pi}\frac{a^{\prime}_s(\lambda)}{1+a_s(\lambda)}=\int_{-\infty}^{+\infty}{\rm d\lambda} \, \rho_s^{\ast} (\lambda)\lambda^2\,,
\ee
where the Bethe equations \eqref{eq:thermo_bethe} have been used. Putting everything together, we obtain the important relation
\be
\frac{{\rm d}}{{\rm d} s}f(s)=
\int_{-\infty}^{+\infty}{\rm d\lambda} \, \rho_s^{\ast}(\lambda)\lambda^2-e_{0}(c)\,. \label{eq:starting_point_appendix_A}
\ee
Accordingly, the small-$s$ behavior of $f(s)$ is determined by that of $\int_{-\infty}^{+\infty}{\rm d\lambda} \, \rho_s^{\ast}(\lambda)\lambda^2$. Note that for $s=0$ we obtain that the derivative of $f(s)$ is the energy of the initial state, as it should. In Appendix \ref{app:small_s_asymptotic} we show that
\be
\int_{-\infty}^{+\infty}{\rm d\lambda} \, \rho_s^{\ast}(\lambda)\lambda^2 = c D^2 -c^2 D^2 \sqrt{\frac{2 s}{\pi}} +O(s)\,,
\label{eq:analytic_prediction}
\ee
so that we finally obtain
\be
f(s) = [c D^2 - e_0(c)]s - \frac{2}{3} c^2 D^2 \sqrt{\frac{2}{\pi}} s^{3/2}+ O(s^2)\,. \label{eq:analytic_low_s}
\ee
We can now plug this expression into Eq.~\eqref{eq:Lengendre_duality} and compute the first term in the expansion of $I(w)$ for $w\simeq \bar{w}^{-}$. By doing so, we obtain that
\be
I(w \rightarrow \bar{w}^{-}) = \frac{\pi}{6 c^4 D^4}(\bar{w} - w)^3 +O((\bar{w} - w)^4)\: \: \, \mbox{for} \: \: w\leq \bar{w}\,,
\label{eq:near_wbar}
\ee
i.e., the first term of the expansion around $\bar{w}$ is cubic instead of quadratic. As anticipated, we therefore find that small fluctuations have not a Gaussian distribution, in stark contrast with the free case.

Next, we proceed to studying the limit of $I(w)$ for small values of $w$. From Eq.~\eqref{eq:Lengendre_duality}, we see that the latter is determined by the behavior of $f(s)$ at $s \rightarrow \infty$, which we now work out analytically. This can be done by following the derivation of Refs.~\cite{Takahashi1973,BePi2018}, where analogous calculations were done in the context of thermal equilibrium. We start by rewriting Eq.~\eqref{eq:final_SCGF} as 
\begin{equation}
	S_{QA}[s,\rho_s^{\ast}]=D(\mbox{ln}\gamma +1) +\frac{h D}{2} -P(s)\,,
\end{equation}
where
\begin{equation}
	P(s)=  \frac{1}{2} \int_{-\infty}^{+\infty}\frac{{\rm d} \lambda}{2 \pi} \mbox{ln}(1+\mbox{e}^{-s \varepsilon_s^{\ast}(\lambda)}). \label{pressure}
\end{equation}
For large $s$, the function $\varepsilon_s^{\ast}(\lambda)$ has two symmetric zeros which we call $Q'$ ($-Q'$), while we name $Q$ ($-Q$) the zeros of $\varepsilon_{\infty}(\lambda)$, defined as the solution of the limit $s \rightarrow \infty$ of Eq.~\eqref{eq:saddlepoint_density}, namely by
\be
\varepsilon_{\infty}(\lambda)=2\lambda^2-h'+\int_{-Q}^{Q}\frac{\rm d\mu}{2\pi}K(\lambda-\mu)\varepsilon_{\infty}(\mu)\,,
\label{eq:lowtemperatureTBA}
\ee
where we made the assumption that 
\begin{equation} 
	0< h' =\lim_{s\to\infty} \frac{1}{s}h(s) <\infty\,. \label{lagrange_multiplier_slope}
\end{equation}
Assuming the validity of the latter equation we write the following expansion for $h(s)$ at large $s$ 
\begin{equation}
	h(s) = h' s + h_0 +\frac{h_{-1}}{s} +O(s^{-2})\,. \label{lagrange_multiplier_line}
\end{equation}
We computed $h'$, $h_0$ and $h_{-1}$ by performing a fit against the numerical data for $h(s)$. The expansion in Eq.~\eqref{lagrange_multiplier_line} has been numerically checked. Next, we write Eq.~\eqref{pressure} as 
\begin{equation}
	\begin{split}
		P(s) &= \frac{1}{2}\int_{-\infty}^{+\infty} \frac{{\rm d} \lambda}{2 \pi} \, \mbox{ln}\left(1+\mbox{e}^{-s|\varepsilon_s^{\ast}(\lambda)|}   \right) - \frac{s}{2}\int_{-Q'}^{-Q} \frac{{\rm d} \lambda}{2 \pi} \varepsilon_s^{\ast}(\lambda) \\ 
		&- \frac{s}{2}\int_{Q}^{Q'} \frac{{\rm d} \lambda}{2 \pi} \varepsilon_s^{\ast}(\lambda) - \frac{s}{2}\int_{-Q}^{Q} \frac{{\rm d} \lambda}{2 \pi} \varepsilon_s^{\ast}(\lambda). 
	\end{split}
	\label{pressuredecomposition} 
\end{equation}
The first term in Eq.~\eqref{pressuredecomposition} can be studied by expanding the integrand around the points $Q'(-Q')$
\begin{equation}
	\frac{1}{2} \int_{-\infty}^{+\infty} \frac{{\rm d} \lambda}{2 \pi} \, \mbox{ln}\left(1+\mbox{e}^{-s|\varepsilon_s^{\ast}(\lambda)|} \right)=\frac{\pi}{12 |\varepsilon_s^{\ast \prime}(Q')|s} + O(s^{-2}). \label{Qintegralexpansions}
\end{equation}
Note that the second and third term in Eq.~\eqref{pressuredecomposition} vanish as $O\left((Q-Q')^2\right)$, i.e.,
\be
\begin{split}
	\int_{Q}^{Q'} \frac{{\rm d} \lambda}{2 \pi} \varepsilon_s^{\ast}(\lambda) &= -\int_{Q'}^{Q'+(Q-Q')} \frac{{\rm d} \lambda}{2 \pi} \varepsilon_s^{\ast}(\lambda) \\  
	&= -\frac{1}{4 \pi} (Q-Q')^2 \varepsilon_s^{\ast \prime}(Q') + O\left( (Q-Q')^3 \right).  \label{Qintegralexpansions_2} 
\end{split}
\ee
We now make use of the following identities, which are proven in Appendix~\ref{app:large_s_asymptotic}:
\begin{align}
	\delta \varepsilon_s^{\ast}(\lambda) = \varepsilon_s^{\ast}(\lambda)-\varepsilon_{\infty}(\lambda)=& \frac{U_1(\lambda)}{s} + \frac{U_2(\lambda)}{s^2} \nonumber \\ 
	&+ O(s^{-3}), \label{eq:expansionLowT_1} \\
	Q'-Q = -\frac{U_1(Q)}{s \varepsilon'_{\infty}(Q)} &+ O(s^{-2})\, \label{eq:expansionLowT_2}
\end{align}
where $U_1(\lambda)$ and $U_2(\lambda)$ are obtained as the solution to the following integral equations:
\begin{align}
	U_1(\lambda) = -h_0 +\mbox{ln}\left[\frac{\lambda^2}{c^2}\left(\frac{\lambda^2}{c^2} +\frac{1}{4}\right)\right] \hspace{2.6cm}
	\nonumber \\ 
	+ \int_{-Q}^{Q}\frac{{\rm d} \mu}{2 \pi} K(\lambda -\mu) U_1(\mu)\,, \label{eq:expansionLowT_3} \\
	U_2(\lambda) = \frac{\left[K(\lambda-Q)+K(\lambda+Q)\right]}{\varepsilon_{\infty}'(Q)}\left(-\frac{U_1^2(Q)}{4 \pi}-\frac{\pi}{12}\right) \nonumber \\
	-h_{-1} + \int_{-Q}^{Q}\frac{{\rm d} \mu}{2 \pi} K(\lambda-\mu)U_2(\mu)\,. \label{eq:expansionLowT_4}
\end{align}
Plugging the identities in Eqs.~\eqref{eq:expansionLowT_1}--\eqref{eq:expansionLowT_4} into Eqs.~\eqref{Qintegralexpansions} and \eqref{Qintegralexpansions_2} and then into Eq.~\eqref{pressuredecomposition}, straightforward manipulations finally yield, for $s \rightarrow \infty$,
\be
f(s)= 2 f_0 + \frac{f_1}{s} +O(s^{-2}),
\label{finallargesasymptoticcomplete}
\ee
with 
\begin{eqnarray}
	f_0 &=& \frac{D}{2}\left(\mbox{ln} \gamma +1\right) +\frac{1}{4}\int_{-Q}^{Q} {\rm d} \lambda \, \rho_{GS} (\lambda) \, \mbox{ln}\left[\frac{\lambda^2}{c^2}\left(\frac{\lambda^2}{c^2}+\frac{1}{4} \right)\right], \nonumber \\
	f_1 &=&  -\frac{1}{v_s}\left(\frac{U_1^2(Q)}{4 \pi}+\frac{\pi}{12} \right).
	\label{finallargesasymptotic}
\end{eqnarray}
Here $v_s$ is the sound velocity of the system defined by
\begin{equation}
	v_s =\frac{\varepsilon'_{\infty}(Q)}{ 2\pi \rho_{GS}(Q)}.
\end{equation}
The expression of $f_0$ in Eq.~\eqref{finallargesasymptotic} coincides with the prediction of Ref.~\cite{GaSi12}, since $f_0$ can be actually rewritten as
\begin{equation}
	f_0=-\frac{\ln|\langle {\rm BEC}|\psi_0^c\rangle|}{L}\,,
	\label{eq:fs_gapped}
\end{equation}
where $|\psi_0^c\rangle$ is the ground state of the post-quench Hamiltonian $H(c)$ in Eq.~\eqref{eq:hamiltonian}.

\begin{figure}
	\centering
	\includegraphics[width=1\columnwidth]{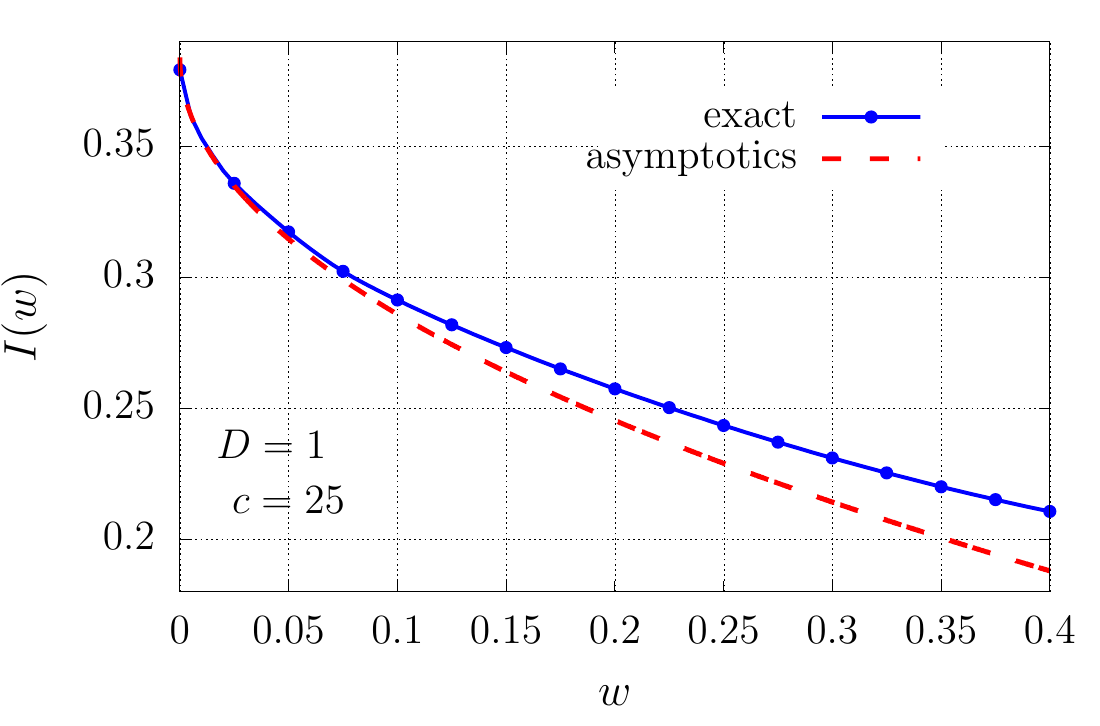}	
	\caption{Rate function $I(w)$ after quenching the interaction parameter to $c=25$. The density of the particles is fixed to $D=1$. The blue solid and red dashed lines correspond, respectively, to the exact numerical value of $I(w)$ [obtained using Eq.~\eqref{eq:proved_Gartner}], and to the analytic expansion in Eq.~\eqref{eq:low_w_I}.}
	\label{fig:largedeviationsLLloww}
\end{figure}

The expression in Eq.~\eqref{finallargesasymptoticcomplete} finally provides access to the behavior of $I(w)$ for small values of $w$. Indeed, by plugging Eq.~\eqref{finallargesasymptoticcomplete} into Eq.~\eqref{eq:Lengendre_duality} we obtain
\be
I(w)= 2 f_0 - 2 \sqrt{-f_1} w^{1/2}+O(w)\,.
\label{eq:low_w_I}
\ee 
We compare this expansion with the exact rate function $I(w)$ obtained by the numerical evaluation of the formulas derived in the previous section for several values of the interaction $c$. An example is displayed in Fig.~\ref{fig:largedeviationsLLloww}, where a good agreement between the two curves is manifest. Note that the leading behavior in the expansion in Eq.~\eqref{eq:low_w_I} is in agreement with the predictions of Ref.~\cite{GaSi12}. Indeed, based on a quantum-to-classical correspondence, it was argued in Ref.~\cite{GaSi12} that when the post-quench Hamiltonian is critical, the rate function displays the generic behavior
\be
I(w)-2f_0\propto w^{d/(d+1)}\,,
\ee
where $d$ is the dimensionality of the system. In our case, the criticality condition is verified since the Lieb-Liniger spectrum is gapless. Accordingly, large deviations for small values of the work encode signatures of universality as predicted by the quantum-to-classical correspondence \cite{GaSi12}. A precise determination of the classical counterpart of the quantum quench analyzed here, however, goes beyond the scope of the present paper, and will not be discussed further.

\subsection{The Tonks-Girardeau limit}
\label{sec:c_infinity_limit}

In this section we focus on the Tonks-Girardeau limit \cite{Gira60}, corresponding to the quench where the final interactions are taken to be infinitely large. In fact, on the one hand, in this regime the formulas derived in the previous sections simplify, so that one can push the analytical control even further. On the other hand, in this limit, qualitative differences emerge in the statistics of the work, which are worth exploring per se, especially given the great relevance of this regime for cold-atomic experiments \cite{KiWW04,HGMD09}.

\begin{figure}
	\centering
	\includegraphics[width=1\columnwidth]{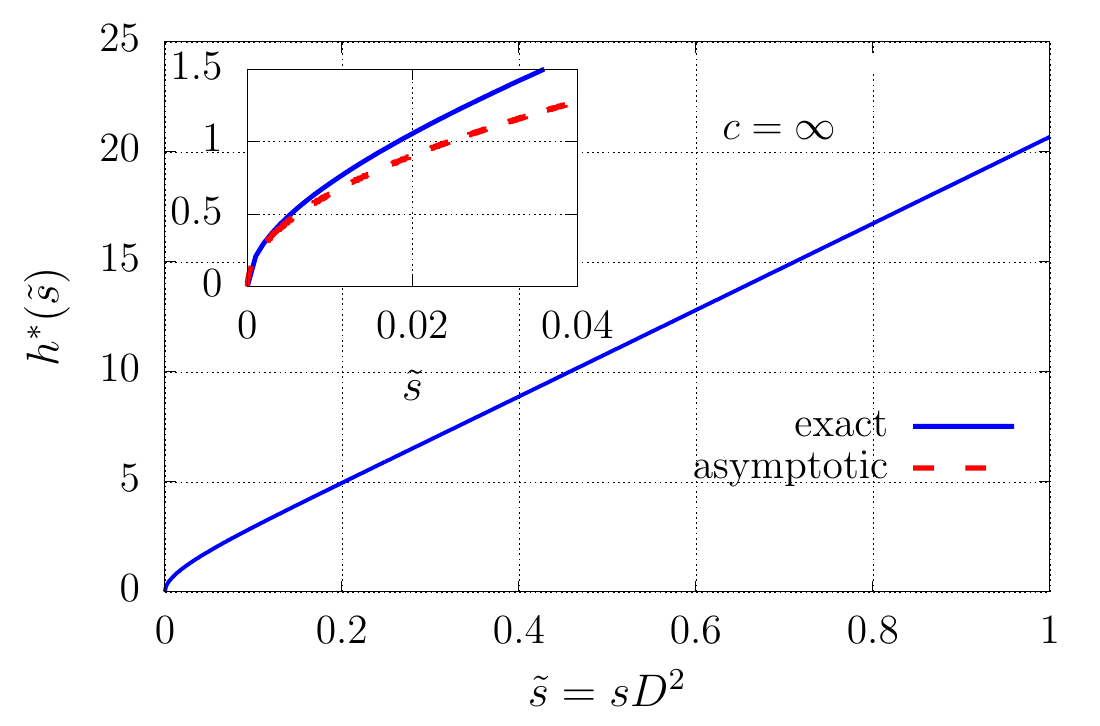}	
	\caption{Lagrange multiplier $h^\ast (\tilde{s})$ as a function of $\tilde{s}=s D^2$. We note that $h^\ast(\tilde{s})$ is nearly linear, except for small values of $\tilde{s}$. Inset: behavior of $h^{\ast}(\tilde{s})$ close to $\tilde{s}=0$. Together with the exact curve $h^\ast (\tilde{s})$ (blue solid line) we report the analytic expansion in Eq.~\eqref{lowshast} (red dashed line).}
	\label{fig:starmultiplier}
\end{figure}

From the computational point of view, in the limit $c \rightarrow \infty$, the kernel $K(\lambda)$ in Eq.~\eqref{eq:k_function}, entering the integral equations which characterize the quench action formalism, vanishes identically, largely simplifying the analysis. In particular, from Eq.~\eqref{eq:saddlepoint_density}, the solution for $\eta_s^{\ast}(\lambda)$ can be explicitly written as
\be
\eta_s^{\ast}(\lambda) =   \frac{\lambda^2}{4 c^2} \mbox{e}^{2s \lambda^2 -h(s)} =  \frac{\lambda^2}{4 D^2}\, e^{2 s \lambda^2 -h^{\ast}(s,D)}\,.
\label{eq:solutionTG}
\ee
Here we have introduced the following parametrization of the Lagrange multiplier $h(s)$
\be
h(s) = h^{\ast}(s,D) -\mbox{ln}(c^2) +\mbox{ln}(D^2)\,,
\label{eq:def_h_ast}
\ee
which is particularly convenient because, as shown in Ref.~\cite{DWBC14}, for the quench action equations corresponding to $s=0$ in the present discussion,  
\be
h(0) =  -\mbox{ln}(c^2) +\mbox{ln}(D^2)\,,
\label{eq:lag_multIBEC}
\ee
and therefore $h^{\ast}(0,D)=0$. 

Next, from Eq.~\eqref{eq:thermo_bethe} one finds, in the TG limit, $\rho^t(\lambda)= 1/(2\pi)$ and thus from $\rho^t(\lambda) =\rho(\lambda)(1+\eta(\lambda))$
\be
\rho_s^{\ast}(\lambda) = \frac{1}{2 \pi} \frac{1}{1+\frac{\lambda^2}{4 D^2} \, e^{2 s \lambda^2 -h^{\ast}(s,D)}}\,.
\label{eq:TG_rhos}
\ee
Accordingly, the density constraint in Eq.~\eqref{eq:density_thermo}, which determines the parameter $h^{\ast}(s,D)$, can be written as
\be 
\int_{-\infty}^{+\infty} \frac{{\rm d} y}{2 \pi} \frac{1}{1+ \frac{y^2}{4} \, \, \mbox{e}^{2 y^2 D^2 s -h^{\ast}(s,D)}} = 1\,.
\label{starmultiplierequation}
\ee 

\begin{figure}
	\centering
	\includegraphics[width=1\columnwidth]{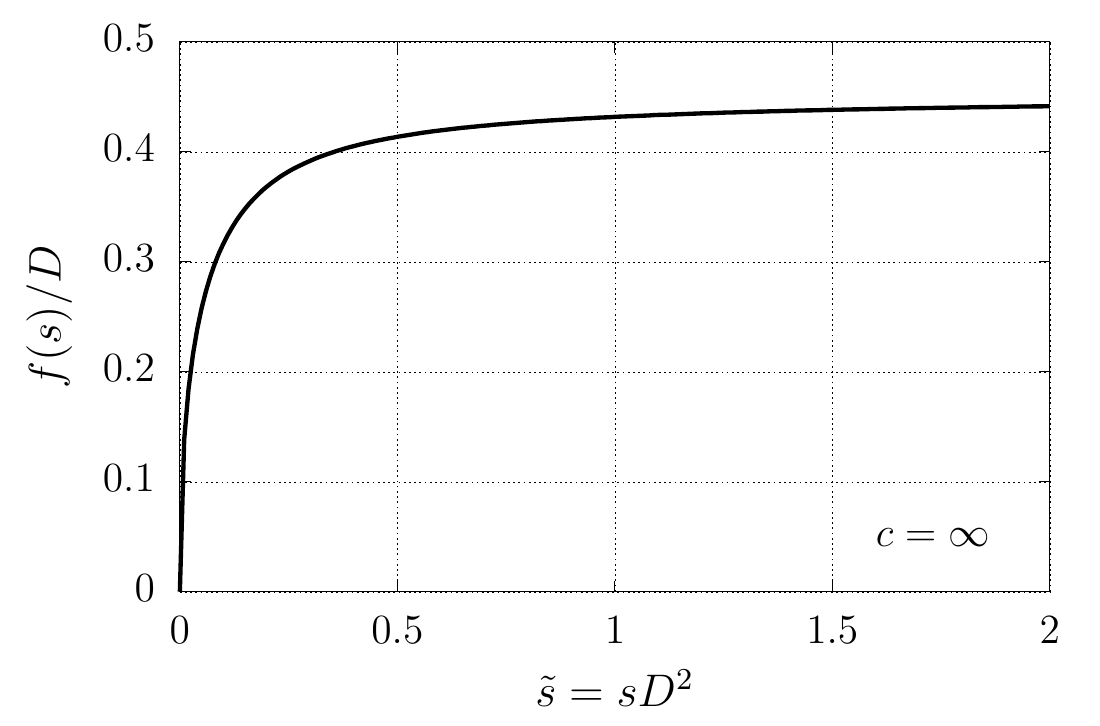}	
	\caption{Scaled cumulant generating function divided by the density of Bosons $f(s)/D $ as a function of $\tilde{s}=s D^2>0$ for quenches to infinitely repulsive interactions $c \rightarrow \infty$ (Tonks-Girardeau limit). The function approaches the origin $\tilde{s} \rightarrow 0$ with infinite slope, according to Eq.~\eqref{lowsasym}.}
	\label{fig:SCGFTG}
\end{figure}

The function $h^{\ast}(s,D)$ determined by this condition does not depend on $c$, since the latter does not appear in Eq.~\eqref{starmultiplierequation}, and it actually depends on $s$ and $D$ via the combination $\tilde{s} = s D^2$. Equation \eqref{starmultiplierequation} can be easily solved numerically: we report the corresponding result for $h^{\ast}(\tilde{s})$ in Fig.~\ref{fig:starmultiplier}. Interestingly, the function $h^{\ast}(\tilde{s})$ appears to be almost linear in $\tilde{s}$. In fact, it is not difficult to compute the asymptotic behavior of $h^{\ast}(\tilde{s})$ for $\tilde{s} \to \infty$. Inserting Eq.~\eqref{lagrange_multiplier_line} written up to order $O(s^{-1})$, $h_0'= h_0 -h(0)$ with $h_0$ and $h(0)$ defined in Eqs.~\eqref{lagrange_multiplier_line} and Eq.~\eqref{eq:lag_multIBEC}, respectively,  
\begin{equation}
	h^{\ast}(\tilde{s}) = \tilde{s} h' + h_0' + O(\tilde{s}^{-1})\,
	\label{largesmultiplierasymptotic}
\end{equation}
into Eq.~\eqref{eq:TG_rhos}, in the large-$s$ limit $\rho^\ast_s(\lambda)$ becomes a step function; in particular, imposing the correct density, we find
\begin{equation}
	\lim_{s \rightarrow \infty} \rho_s^{\ast}(\lambda)  = \left\{
	\begin{array}{lr}
		\frac{1}{2 \pi} \; \; |\lambda|<D \pi \\[2mm] 
		\; 0   \; \; \, \, |\lambda|>D \pi
	\end{array}
	\right.
\end{equation}
and therefore the simple relation $h'=2 \pi^2 $ in the TG limit. The behavior of $h^\ast (\tilde{s})$ can be analytically studied also for $\tilde{s} \rightarrow 0$, although the computations are more involved. For this reason, we present them in Appendix \ref{app:large_s_asymptotic_TG}, and we report here the final result, which reads
\begin{equation}
	h^{\ast}(\tilde{s}) = \frac{16}{\sqrt{2 \pi}} \sqrt{\tilde{s}} +O(\tilde{s})\,.
	\label{lowshast}
\end{equation}

The expressions in Eqs.~\eqref{largesmultiplierasymptotic} and \eqref{lowshast} allow us to obtain directly the corresponding expansions for the scaled cumulant generating function $f(s)$. First, note that in the TG limit the latter can be expressed explicitly by plugging Eq.~\eqref{eq:solutionTG} into Eqs.~\eqref{eq:proved_SCGF} and \eqref{eq:final_SCGF}, finding 
\begin{equation}
	\begin{split}
		f(s)=D &+ \frac{D}{2}h^{\ast}(s D^2) -\frac{D^3 \pi^2}{3}s \\ 
		&-\frac{1}{2} \int_{-\infty}^{\infty} \frac{\mbox{d} \lambda}{2 \pi} \, \mbox{ln}\left(1+ \frac{4D^2}{\lambda^2}\mbox{e}^{-2 s\lambda^2+ h^\ast(s,D)} \right)\,.
		\label{finalresultTG}
	\end{split}
\end{equation}
This expression can be easily evaluated numerically, as it amounts to a simple integral, once the function $h^{\ast}(s D^2)$ is known. Notice that, rescaling $\lambda = D y$ in the integral of Eq.~\eqref{finalresultTG}, it turns out that $f(s)/D$ is actually a function of $\tilde{s}=s D^2$ only, as it has been already noticed also for $h^{\ast}(s,D^2)$. We report the resulting data for $f(s)/D$ in Fig.~\ref{fig:SCGFTG}. The same rescaling in terms of $D$ does not apply to the case at finite $c$ of Secs.~\ref{sec:numerical_results} and \ref{sec:asymptotics_calculations} as one realizes by looking, e.g., at the small-$s$ expansion in Eq.~\eqref{eq:analytic_low_s}. Note also that, differently from the case of finite interactions $c$, the average value of the work is infinite in the Tonks-Girardeau limit. Indeed, the expansion of $f(s)/D$ near $s =0$ differs from Eq.~\eqref{eq:analytic_low_s}, as one realizes by plugging Eq.~\eqref{lowshast} into Eq.~\eqref{finalresultTG}. The details of this calculation are reported in Appendix~\ref{app:low_s_asymptotic_TG}, where we find
\be
\frac{f(s)}{D} = \alpha_{1/2} \; \tilde{s}^{1/2} + O\left(\tilde{s} \right)\,,
\label{lowsasym}
\ee
with
\begin{eqnarray}
	\alpha_{1/2} &=& 2 \sqrt{\frac{2}{\pi}}\,.
\end{eqnarray} 

Analogously, the asymptotic behavior of $f(s)$ for $s \rightarrow \infty$ can be derived by plugging in Eq.~\eqref{finalresultTG}  the expression in Eq.~\eqref{largesmultiplierasymptotic} for $h^{\ast}(\tilde{s})$. The intermediate steps are reported in Appendix~\ref{app:large_s_asymptotic_TG}, while the final result reads
\be
\frac{f(s)}{D} = 2 f_0 + \frac{f_1}{\tilde{s}} + O(\tilde{s}^{-2})\,, 
\label{eq:large_s_asympt_TG}
\ee
with
\bea
f_0 &=& \frac{1}{2} \, {\rm ln}\; \frac{\pi}{2}\, \label{largesasym0}, \\
f_1 &=&  \frac{1}{8 \pi^2} \left[ -\frac{\pi^2}{6} -\frac{1}{2} \mbox{ln}^2\left(\frac{\pi^2}{4} \mbox{e}^{-h^{\prime}_0} \right) \right]\,.  \label{largesasym}
\eea
Note that the expression for $f_1$ in Eq.~\eqref{largesasym} is equal to the limit $c \rightarrow \infty$ of Eq.~\eqref{finallargesasymptotic}, as it should.

\begin{figure}
	\centering
	\includegraphics[width=1\columnwidth]{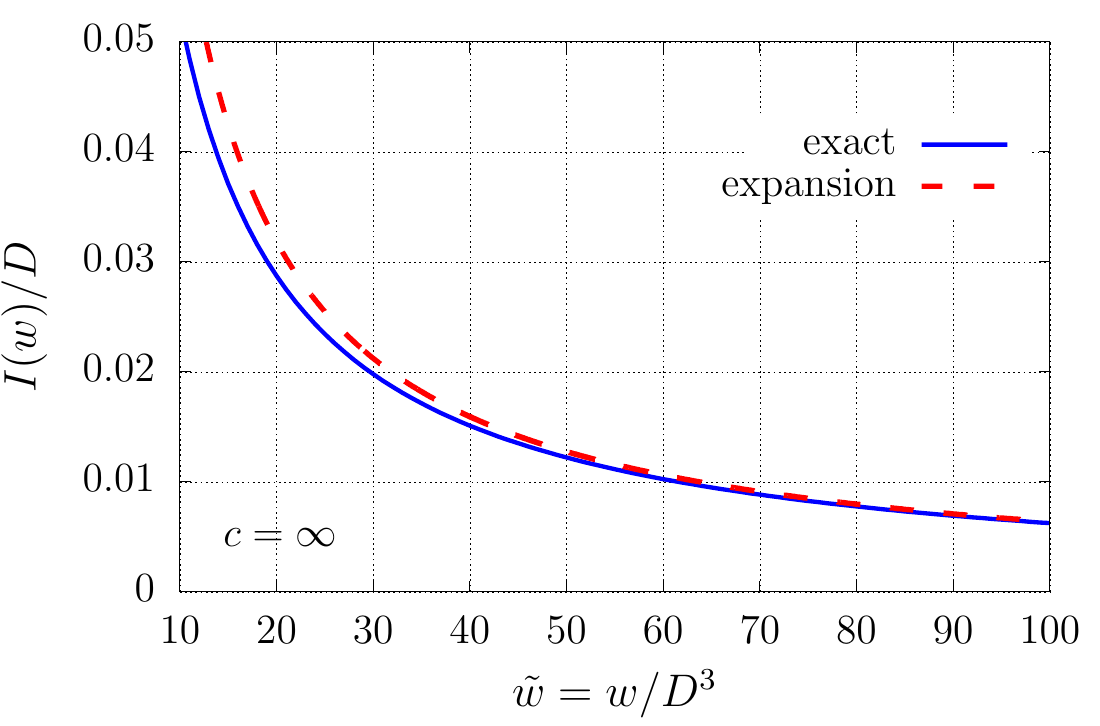}	
	\caption{Asymptotic behavior of the rate function $I(w)/D$ for large values of $\tilde w = w/D^3$ and quenches to the TG regime $c \rightarrow \infty$. The solid line corresponds to the exact value obtained by the numerical Legendre-Fenchel transform of the function $f(s)/D$ given in Eq.~\eqref{finalresultTG}, while the dashed line is the analytical expansion in Eq.~\eqref{largehyperbola}.}
	\label{fig:highw}
\end{figure}

Finally, we discuss the rate function $I(w)$ in the TG limit which can be computed by numerically performing the Legendre-Fenchel transform of Eq.~\eqref{finalresultTG}, displayed in Fig.~\ref{fig:compare_interactions_f_I}(b). Given the scaling form of $f(s)/D$ as a function of $s D^2$, one readily obtains from Eq.~\eqref{eq:proved_Gartner} a scaling form for $I(w)/D$ as a function of the variable $\tilde{w}=w/D^3$ only. Contrary to the case of finite interactions, $I(w)/D$ never vanishes, as the average work $\bar{w}$ grows to infinity as $c\to \infty$. It is thus meaningful to study the asymptotic behavior of $I(w)/D$ for large values of $w/D^3$. This can be easily done by plugging into Eq.~\eqref{eq:Lengendre_duality} the expansion in Eq.~\eqref{lowsasym}, finding
\be
\frac{I(w)}{D} = \frac{\alpha_{1/2}^2}{4} \tilde{w}^{-1} +O\left(\tilde{w}^{-2}\right)\,,
\label{largehyperbola}
\ee
which is plotted in Fig.~\ref{fig:highw} together with the exact numerical values of $I(w)/D$.

Before concluding this section, we note that an analogous analysis can be done for the limit $w \rightarrow 0$ of the rate function $I(w)$, by plugging Eq.~\eqref{eq:large_s_asympt_TG} into Eq.~\eqref{eq:Lengendre_duality}. In this case, we find that $I(w)$ behaves as in Eq.~\eqref{eq:low_w_I}, showing that fluctuations for small values of $w$ are not qualitatively affected by considering the TG regime $c \rightarrow \infty$.

\section{Algebraic behavior of $p(w)$ for large $w$}
\label{sec:power_law_tail}

In the previous sections we have quantitatively analyzed the rate function $I(w)$, characterizing the exponential decay of the distribution function $p(w)$ for $w<\bar{w}$ as $L$ grows. On the other hand, for the quench considered here, we saw that $I(w)$ vanishes identically for $w>\bar{w}$, so that, in this regime, the decay of $p(w)$ as a function of $L$ is sub-exponential. As we have anticipated in Sec.~\ref{sec:exact_rate_functions}, assuming (for large $w$) a power-law decay $p(w)\sim C(L,c) \, w^{-\beta}$ one can bound the value  of $\beta$ based on the divergence of cumulants of $w$ higher than the first. However, the leading behavior in $L$ of $C(L,c)$, characterizing $p(w)$ for $w>\bar{w}$, and the value of the exponent $\beta$, can not be obtained by large deviation theory and a more sophisticated analysis has to be carried out.

In order to determine such a leading behavior, one could start from an exact expression for $p(w)$ at finite sizes, and then perform the correct asymptotic analysis. This strategy, however, appears to be unpractical, as the exact computation of $p(w)$ at finite sizes is a significant challenge. Nevertheless, in this section we show that this problem can be solved assuming a vanishingly small density of particles. In particular, we consider the limit $L\to \infty$, with the number $N$ of particles kept fixed. We will show that, in this regime, one can extract the exact leading dependence on $L$ of $p(w)$ for large values of $w$.

From the technical point of view, the reason why the problem becomes tractable in this regime lies in the simplified structure of the solution to the Bethe equations \eqref{eq:bethe_eq}, parametrizing the eigenstates of the Hamiltonian. Indeed, fixing the quantum numbers $I_j$ in Eq.~\eqref{eq:bethe_equations_log}, one immediately obtains the following expansion as $L$ grows
\begin{equation}
	\lambda_j = \frac{2 \pi I_j}{L} +\sum_{n=2}^{\infty}\frac{\lambda_j^{(n)}}{L^n}= \frac{2 \pi I_j}{L}+O(L^{-2})\,,
	\label{eq:smalldensityrapidities}
\end{equation}
with $j=1\,,\ldots N$. Namely, at the leading order in $L$, the rapidities coincide with the quasi-momenta of a free quantum gas confined within a ring of length $L$ (with the condition $I_j\neq I_k$ for $j\neq k$). The corresponding eigenvalues of the Hamiltonian become
\begin{equation}
	E_{N}= \sum_{j=1}^{N} \lambda_j^2 = \frac{4 \pi^2}{L^2} \sum_{j=1}^{N} I_j^2 +O(L^{-3})\,. \label{eq:small_density_eigenvalues}
\end{equation}
For simplicity, we consider below the case in which the number $N$ of particles is odd, so that the quantum numbers $I_j$ are integer, i.e., we choose
\be
N=2M+1\,.
\ee
A completely analogous derivation applies to the case of even $N$.

As a first ingredient for the computation of $p(w)$, we consider the zero-density limit of the overlap in Eq.~\eqref{eq:overlaps}, which has been already studied in the literature \cite{BDWC14, DoCa12,CaLe14}. In particular one finds
\be 
\langle  \{\lambda_j\} | \mathrm{BEC} \rangle \simeq \frac{\sqrt{(cL)^{-M}} \sqrt{N!}}{\prod_{j=1}^{M}\left(\frac{\lambda_j}{c} \sqrt{\frac{1}{4}+\frac{\lambda_j^2}{c^2}}\right)}\,. \label{smalldensityoverlaps}
\ee 
As a consistency check, one should verify that, using Eq.~\eqref{smalldensityoverlaps} and keeping only the leading term of the rapidities $\lambda_j\simeq 2\pi I_j /L$, one obtains the correct values for the normalization and the energy of the initial state, i.e.,
\be
\sum_{I_1<I_2<...I_M} |\langle  \{\lambda_j\} | {\rm BEC} \rangle|^2 =1\,,
\label{eq:overlapschecks}
\ee
and
\be
\sum_{I_1<I_2<...I_M}  \left(\sum_{j=1}^{M} 2 \lambda_j^2 \right) |\langle  \{\lambda_j\} | {\rm BEC} \rangle|^2 = \frac{cN(N-1)}{L}\,. 
\label{eq:energychecks}
\ee
In fact, Eq.~\eqref{eq:overlapschecks} can be established analytically on the basis of Eq.~\eqref{smalldensityoverlaps} by using the identity (see, e.g., Ref.~\cite{Zagier2012})
\begin{equation}
	\sum_{1\leq I_1<I_2<...I_M<\infty} \prod_{j=1}^{M} \frac{1}{I_j^2}  = \frac{\pi^{2M}}{(2M+1)!}\,,
\end{equation}
while we checked that also Eq.~\eqref{eq:energychecks} is fulfilled by numerically performing the sum for small particle numbers.

We have now all the ingredients to determine the leading behavior of $p(w)$. Our strategy consists in a direct computation based on the definition in Eq.~\eqref{eq:work_probability_definition} which, for the Lieb-Liniger model, reads
\begin{equation}
	P(W) = \sum_{\{\lambda_j \}} |\langle \{\lambda_j\} | {\rm BEC} \rangle|^2 \delta \left( W-(E[\{\lambda_j \}] -E_0^c) \right)\,.
	\label{probabilitydefinition}
\end{equation}
Let us fix the value $w$, so that the extensive work is $W=wL$, and let $\epsilon\ll W$ be a small energy shell (more precisely, we choose $\epsilon=\tilde{\epsilon}L$ with $\tilde{\epsilon}\ll w$). Then, the definition in Eq.~\eqref{probabilitydefinition} directly yields
\begin{equation}
	\sum_{W' \in (W-\epsilon,W+\epsilon)} P(W') \sim \epsilon \, P(W)\,. 
	\label{workshelldefinition}
\end{equation}
Note that since $w=O(L^{0})$, we are in the regime $w\gg \bar{w}$, since in the zero-density limit $\bar{w}=O(L^{-2})$. We can then proceed to evaluate the sum over the energy shell in Eq.~\eqref{workshelldefinition} and obtain the behavior of $P(W)$ and hence of $p(w)$. In order to simplify the discussion, we start by illustrating the main idea of this derivation in the simplest case where $M=1$, i.e., $N=3$. The generalization to an arbitrary number $N$ of particles, which does not bear conceptual complications, is presented in Appendix \ref{app:power_law}. At the end of this section we will report the final result of this analysis.

For $M=1$, Eq.~\eqref{workshelldefinition} can be rewritten as
\begin{equation}
	\epsilon P(W) = \sum_{I_1 \in (I_{\rm min},I_{ \rm max})} |\langle I_1 | \mathrm{BEC} \rangle|^2 \,,
\end{equation}
where $I_{\rm min}$ and $I_{ \rm max}$ are determined, via Eq.~\eqref{eq:small_density_eigenvalues}, by the boundaries of the energy shell in Eq.~\eqref{workshelldefinition}, i.e.,
\begin{eqnarray}
	I_{\rm min} &=& \frac{L}{\sqrt{8} \pi} \sqrt{W-\epsilon}\,, \label{shellIvalues_min}\\
	I_{\rm  max} &=& \frac{L}{\sqrt{8} \pi} \sqrt{W+\epsilon}\,. \label{shellIvalues_max} 
\end{eqnarray}
Here we dropped the ground state energy $E_0^c \sim 1/L^2$ since it is sub-leading with respect to $W$. Consequently we have
\begin{equation}
	\begin{split}
		\epsilon P(W) &= \frac{6 c^2 L^2}{16 \pi^2} \sum_{I=I_{\rm min}}^{I_{\rm max}} \frac{1}{I_1^2 (\pi^2 I_1^2 + \frac{c^2 L^2}{16})} \\
		&= \frac{3 c^2 L^2}{8 \pi^2} \sum_{j=0}^{\delta I} \frac{1}{(j+I_{\rm min})^2 (\pi^2 (j+I_{\rm  min})^2 +\frac{c^2 L^2}{16} )}, \label{intermediateresultlowdensity}
	\end{split}
\end{equation} 
where
\be
\delta I \equiv I_{\rm  max}-I_{\rm min} = \frac{L \epsilon}{\sqrt{8 W} \pi} + O\left( \epsilon^2 \right)\,. \label{shellIvalues}
\ee
The series in Eq.~\eqref{intermediateresultlowdensity} can be easily bounded as
\be
\epsilon P_{\rm min} < \epsilon P(W)< \epsilon P_{\rm max}\,,
\ee
where
\bea
\epsilon P_{\rm min} &=& \frac{3 c^2 L^2}{8 \pi^2} \sum_{j=0}^{\delta I} \frac{1}{I_{\rm max}^2 (\pi^2 I_{\rm max}^2 + \frac{c^2 L^2}{16})} \nonumber\\ 
&=& \frac{3 c^2 L^2}{8 \pi^2} \frac{\delta I}{I_{\rm max}^2 (\pi^2 I_{\rm max}^2 + \frac{c^2 L^2}{16})}\,, \label{pminbounds}\\
\epsilon P_{\rm max} &=& \frac{3 c^2 L^2}{8 \pi^2} \sum_{j=0}^{\delta I} \frac{1}{I_{\rm min}^2 (\pi^2 I_{\rm min}^2 + \frac{c^2 L^2}{16})}\nonumber\\
&=& \frac{3 c^2 L^2}{8 \pi^2} \frac{\delta I}{I_{\rm min}^2 (\pi^2 I_{\rm min}^2 + \frac{c^2 L^2}{16})}\,. \label{pmaxbounds}
\eea
Plugging Eqs.~\eqref{shellIvalues_min}--\eqref{shellIvalues} into Eqs.~\eqref{pminbounds} and \eqref{pmaxbounds}, we immediately get
\bea
P_{\rm min} &=& \frac{3 c^2 L^3}{ \pi^3 \sqrt{8^3W}}  \frac{1}{W(\frac{c^2 L^4}{128 \pi^2}+\frac{W L^4}{64 \pi^2})}+O(\epsilon)\,,\\
P_{\rm max} &=&\frac{3 c^2 L^3}{ \pi^3 \sqrt{8^3W}} \frac{1}{W(\frac{c^2 L^4}{128 \pi^2}+\frac{W L^4}{64 \pi^2})}+O(\epsilon)\,. \label{eq:P_bound_1_particles}
\eea
Accordingly, at the leading order in $\epsilon$, we find $P_{\rm min}=P_{\rm max}=P(W)$. We recall now that the probability distribution function $p(w)$ of the intensive work is related to $P(W)$ by $p(w)=L \, P(w L)$. Accordingly, from Eq.~\eqref{eq:P_bound_1_particles}, we obtain 
\be  
p(w) \propto  \frac{c^2}{L^{5/2}} \, w^{-5/2} + O(L^{-7/2})\,.
\label{onerapiditydistribution_intensive}
\ee 
which has the form anticipated in Eq.~\eqref{powerlawtail} with $C(L,c)=c^2/L^{5/2}$, $\beta=5/2$ and clearly displays that the dependence $C(L,c)$ of $p(w)$ on $L$ is \textit{not} exponential. The value of $\beta$ furthermore satisfies the bound $2<\beta <3$, anticipated in Sec.~\ref{sec:numerical_results}. Finally, we show in Appendix~\ref{app:power_law}  that the same reasoning can be repeated for an arbitrary number $N$ of particles and that the behavior in Eq.~\eqref{onerapiditydistribution_intensive} is valid for all $N$ (with an $N$-dependent numerical prefactor).  To our knowledge, this constitutes the first quantitative calculation of the power-law tail of $p(w)$, first predicted in Ref.~\cite{GaSi12}, for interactive bosonic systems starting from a critical initial state. Note that, even if the result in Eq.~\eqref{onerapiditydistribution_intensive} holds for an arbitrary finite number of particles $N$, its validity is still limited to the regime of vanishing density $D$: indeed, by construction, the large-$L$ limit is taken while $N$ is kept finite. In fact, in the case of finite $D$, while the bound $2<\beta <3$ continues to hold, we can not make any statement on the exact value of the exponent without further assumptions.

\section{Conclusions}
\label{sec:conclusions}

In this work we studied the large-deviation statistics of the intensive work $w$ done by an interaction quench of the one-dimensional Lieb-Liniger model, focusing on the case in which the initial state is the ground state of the non-interacting gas. By means of the quench action approach, we have shown that, for $w<\bar{w}$, the large-deviation principle applies to the probability $p(w)$, as it depends exponentially on the system size $L$ with $p(w) \sim {\rm exp}[-L I(w)]$, and that the G\"{a}rtner-Ellis theorem employed in Ref.~\cite{GaSi12} can be used in order to determine the corresponding rate function $I(w)$. We have provided a fully quantitative analysis of the latter, working out analytically its behavior for small values of the intensive work $w$, cf. Eq.~\eqref{eq:low_w_I}, and close to the average work $\bar{w}$, cf. Eq.~\eqref{eq:near_wbar}. Interestingly, we have shown that for $w\simeq \bar{w}^{-}$ fluctuations are not Gaussian, in contrast to what would be expected from a direct application of the central-limit theorem. Furthermore, we analyzed the probability distribution function of the intensive work $p(w)$ for $w\gg \bar{w}$ where the large-deviation principle is violated and $p(w)$ has a sub-exponential dependence on $L$. Using an exact Bethe ansatz representation of the eigenstates of the Hamiltonian, we have derived the power-law decay of $p(w)$ in the regime of vanishing particle density, see Eq.~\eqref{onerapiditydistribution_intensive}, providing the first quantitative calculation of the power-law tail of $p(w)$ for interactive bosonic systems starting from a critical initial state.

Contrary to other works \cite{SoGS13,PaSo14,RyAn18}, our approach to derive $p(w)$ is not based on the Fourier transform of the so-called Loschmidt echo evaluated at real times. This is an important point, as the latter has proven to be especially hard to compute \cite{PiPV17_II,PiPV18}, due to the presence of points of ``non-analyticity'' arising in its real-time dynamics \cite{HePK13}. For this reason the quench action approach presented in this work can be straightforwardly generalized to a wide class of quantum quenches in other interacting integrable models. A particular interesting example would be the prototypical $XXZ$ Heisenberg chain, where, for instance, the quantum-classical correspondence mentioned in Sec~\ref{sec:asymptotics_calculations} could be investigated in detail.

Finally, it would be interesting to investigate the statistics of the work done by quenches to the attractive regime of the Lieb-Liniger model, where intriguing phenomena, such as the formation of multi-particle bound states  \cite{PiCE16}, have been predicted. While the study of arbitrary attractive interactions might be challenging due to the emergence of singularities in the spectrum of the Hamiltonian \cite{takahashi_book}, we believe the weakly attractive regime investigated in Refs.~\cite{CaCR00,CaCR00_II, KaSU03,FlFP16,PiCa16} to be within the reach of the techniques presented in this work. 

\section{Acknowledgments}
G.P. thanks D.Rossini for fruitful discussions. L.P. acknowledges support from the Alexander von Humboldt foundation and from the Deutsche Forschungsgemeinschaft under Germany's Excellence Strategy – EXC-2111 – 390814868. A.G. acknowledges fruitful discussions with A.Silva and F.H.L.Essler. 

\onecolumngrid


\appendix

\section{Small-$s$ asymptotics of the scaled cumulant generating function}
\label{app:small_s_asymptotic}

In this appendix we derive Eqs.~\eqref{eq:analytic_prediction} and \eqref{eq:analytic_low_s}, characterizing the behavior of the scaled cumulant generating function $f(s)$ as $s \rightarrow 0$. The starting point is Eq.~\eqref{eq:starting_point_appendix_A} in the main text and therefore the expansion in $s$ of $\rho_s^{\ast}(\lambda)$ is needed. The latter can be obtained from the expansions in $s$ of $a_s(\lambda)$ and $\rho_s^{t}(\lambda)$ that we now perform.  

First, from the integral equation of $\eta_s^{\ast}(\lambda)$ we can write that for $a_s(\lambda)$, which reads
\be
\log a_s(\lambda) =- 2s \lambda^2 - \log\left[ \frac{\lambda^2}{c^2}\left( \frac{\lambda^2}{c^2} +\frac{1}{4}    \right)\right]+h(s)+\int_{-\infty}^{\infty}\frac{{\rm d} \mu}{2 \pi} \, K(\lambda-\mu) \, \mbox{ln}\left(1+ a_s(\mu) \right)\,. \label{eq:a_saddlepointdensity}
\ee
As a first step, we write the formal expansions of $\log a_s(\lambda)$ and $h(s)$ as a function of $s$:
\bea
\log a_s(\lambda)&=&\log a^{(0)}(\lambda)+s\log a^{(1)}(\lambda)+\frac{s^2}{2}\log a^{(2)}(\lambda)+\ldots\,,\\
h(s)&=&h_0+sh_1+\frac{s^2}{2}h_2+\ldots\,.
\eea
Plugging these into Eq.~\eqref{eq:a_saddlepointdensity}, we obtain a system of integral equations, one for each successive order in the expansion in $s$. The order $0$ gives the same quench action equations solved in Ref.~\cite{DWBC14}. To clarify the procedure we write in addition the result at first order
\be
\log a^{(1)}(\lambda)=-2\lambda^2+h_1+\int_{-\infty}^{+\infty}\frac{{\rm d}\mu}{2\pi} \, K(\lambda-\mu)\frac{a^{(0)}(\mu)}{1+a^{(0)}(\mu)}\log a^{(1)}(\mu)\,. \label{eq:app_first_order_a_small_s}
\ee
Since, for each fixed value of $s$, the driving term of Eq.~\eqref{eq:a_saddlepointdensity} grows as $\lambda^2$ when $\lambda \rightarrow \infty$, it follows that $\log a_s(\lambda)$ increases at most as $\lambda^2$ in the same limit, and therefore the following expansion as a function of $\lambda$ can be written:
\be
\log a_s(\lambda)-\log a^{(0)}(\lambda)=\beta_2(s)\lambda^2+\beta_0(s)+\beta_{-2}\lambda^{-2}(s)+\ldots \,, \label{eq:log_a_expansion}
\ee
where $\beta_{2j}(s)=O(s)$ and $j \leq 1$ is an integer number. In particular, one has
\bea
\beta_2(s)&=&-2s+O(s^2)\,.
\eea
Accordingly, from Eq.~\eqref{eq:log_a_expansion} one has 
\be
a_s(\lambda)=a_0(\lambda)e^{\beta_2(s)\lambda^2+\beta_0(s)+\beta_2\lambda^{-2}+\ldots}=a_0(\lambda)e^{\beta_2(s)\lambda^2}\sum_{n=0}^\infty\alpha_n(s) \lambda^{-2n}\,. \label{eq:a_expansion}
\ee
Since $a_s(\lambda)$ equals $a_0(\lambda)$ in $s=0$ we have that
\bea
\alpha_n(s) = \delta_{n,0} + O(s) \quad {\rm for} \, \, s \rightarrow 0 \label{eq:app_alpha_coefficients}
\eea 
with $\delta_{n,0}$ the Kronecker delta symbol. Next, since the driving term of Bethe equations \eqref{eq:thermo_bethe} is $1/2\pi$, with a reasoning analogous to the one done to justify Eq.~\eqref{eq:log_a_expansion} one has the asymptotic expansion for $\rho_s^{t}$ for large $\lambda$
\be
\rho_s^{t}(\lambda)=\frac{1}{2\pi}+\sum_{n=1}^{\infty}\gamma_{2n}(s)\lambda^{-2n}, \label{eq:app_rho_t_expansion}
\ee
with suitable coefficients $\gamma_{2n}(s)$ whose explicit expression we do not need for the present calculation.
We can eventually use the results in Eqs.~\eqref{eq:a_expansion} and \eqref{eq:app_rho_t_expansion} into the integral in the r.h.s. of Eq.~\eqref{eq:starting_point_appendix_A}, that we conveniently write as
\be
\int_{-\infty}^{+\infty}{\rm d\lambda} \, \rho_s^{\ast}(\lambda)\lambda^2= 2\int_0^{1}{\rm d\lambda} \, \rho_s^{\ast}(\lambda)\lambda^2+ 2\int_1^{+\infty}{\rm d\lambda} \, \rho_s^{\ast}(\lambda)\lambda^2\,.
\ee
Since the first integral on the r.h.s. has a finite support, one can expand $\rho_s^{\ast}(\lambda)$ in a power series in $s$. Each term can be integrated without divergences, so that after integration the result is expected to have the form of a power series in $s$. Hence, no term proportional to $\sqrt{s}$ will arise from this contribution. This is not the case for the second integral on the r.h.s. as we see in the following. 
First, we write it as
\be
\int_{1}^{+\infty}{\rm d\lambda} \, \rho_s^{\ast}(\lambda)\lambda^2=
\int_{1}^{+\infty}{\rm d\lambda} \, \frac{a_s(\lambda)}{1+a_s(\lambda)}\left(\frac{1}{2\pi}+\sum_{n=1}^{\infty}\gamma_{2n}(s)\lambda^{-2n}\right)\lambda^2\,,
\label{eq:almost_final}
\ee
where we used $\rho(\lambda)=\rho^{t}(\lambda)a(\lambda)/(1+a(\lambda))$ for $\rho^{\ast}_s(\lambda)$ and Eq.~\eqref{eq:app_rho_t_expansion}.  It is not difficult to see that substituting $a_s(\lambda)$ with $a_0(\lambda)$ in the denominator of the integrand on the r.h.s. of Eq.~\eqref{eq:almost_final} leads only to corrections of order $O(s)$. As we are interested in the emergence of terms of order $O(\sqrt{s})$, we are allowed to perform this substitution.  Next we can use Eq.~\eqref{eq:a_expansion} to express $a_s(\lambda)$ in the numerator of the resulting integrand of Eq.~\eqref{eq:almost_final} and then note that \cite{DWBC14}
\be
\frac{a_0(\lambda)}{1+a_0(\lambda)} \simeq \frac{c^2 D^2}{\lambda^4}+O(\lambda^{-6})\,,
\ee
such that Eq.~\eqref{eq:almost_final} becomes
\be
\int_{1}^{+\infty}{\rm d\lambda}\rho_s^{\ast}(\lambda)\lambda^2=
\int_{1}^{+\infty}{\rm d\lambda}\left(\frac{c^2 D^2}{2\pi\lambda^2}+\sum_{n=2}^{\infty}z_{2n}(s)\lambda^{-2n}\right)e^{\beta_2(s)\lambda^2}\,,
\label{eq:final}
\ee
where we used Eq.~\eqref{eq:app_alpha_coefficients} for $\alpha_0$, which again leads to corrections of order $O(s)$, while the coefficients $z_{2n}(s)$ can in principle be derived from $\alpha_{2n}(s)$ and $\gamma_{2n}(s)$ introduced in Eqs.~\eqref{eq:a_expansion} and \eqref{eq:app_rho_t_expansion}, respectively.
The integration in Eq.~\eqref{eq:final} can be performed term by term and only the first one yields a contribution $O(\sqrt{s})$ which is easily computed. Putting the latter result together with Eq.~\eqref{eq:firt_derivative}, which fixes the term linear in $s$, one arrives at Eq.~\eqref{eq:analytic_prediction}, which we also tested numerically. 

\section{Large-$s$ asymptotics of the scaled cumulant generating function}
\label{app:large_s_asymptotic}

In this appendix we provide a detailed derivation of Eqs.~\eqref{eq:expansionLowT_1}--\eqref{eq:expansionLowT_4} reported in the main text.

We start from the equation \eqref{eq:saddlepoint_density} for $\varepsilon_s^{\ast}(\lambda)$ which in the large-$s$ limit reads as 
\begin{equation}
	\varepsilon_s^{\ast}(\lambda)=2\lambda^2-h' -\frac{h_0}{s}-\frac{h_{-1}}{s^2}+\frac{1}{s}\ln\left[\frac{\lambda^2}{c^2}\left(\frac{\lambda^2}{c^2}+\frac{1}{4}\right)\right]-\frac{1}{s}\int_{-\infty}^{\infty}\frac{\rm d\mu}{2\pi}K(\lambda-\mu)\ln\left(1+e^{-s\varepsilon_s^{\ast}(\mu)}\right)+O(s^{-3})\,, \label{firststeplarges}
\end{equation}
where we used the large-$s$ asymptotic of $h(s)$ in Eq.~\eqref{lagrange_multiplier_line}. 
Taking the difference between Eq.~\eqref{firststeplarges} and Eq.~\eqref{eq:lowtemperatureTBA} we have
\begin{equation}
	\varepsilon_s^{\ast}(\lambda)-\varepsilon_{\infty}(\lambda) = -\frac{h_0}{s}-\frac{h_{-1}}{s^2}+\frac{1}{s}\mbox{ln}\left[\frac{\lambda^2}{c^2}\left(\frac{\lambda^2}{c^2} + \frac{1}{4} \right) \right]-\frac{1}{s}\left[\int_{-\infty}^{\infty} \frac{{\rm d} \mu}{2 \pi} K(\lambda -\mu)\left(\mbox{ln}\left(1+\mbox{e}^{-s\varepsilon_s^{\ast}(\mu)} \right) +s\varepsilon_{\infty}^{-}(\mu) \right)\right] +O(s^{-3})\,,
	\label{secondsteplarges}
\end{equation}
where $\varepsilon_{\infty}^{-}(\lambda)$ is defined as 
\be
\varepsilon_{\infty}^{-}(\lambda)= \frac{1}{2} (\varepsilon_{\infty}(\lambda) -|\varepsilon_{\infty}(\lambda)|)\,.
\label{definitionepsilonminus}
\ee
The last integral in Eq.~\eqref{secondsteplarges} can be decomposed in a way similar to Eq.~\eqref{pressuredecomposition}, namely,
\begin{eqnarray}
	\int_{-\infty}^{\infty} \frac{{\rm d} \mu}{2 \pi} K(\lambda -\mu)\left(\mbox{ln}\left(1+\mbox{e}^{-s\varepsilon_s^{\ast}(\mu)} \right) +s\varepsilon_{\infty}^{-}(\mu) \right) &=& \int_{-\infty}^{+\infty} \frac{{\rm d} \mu}{2 \pi} K(\lambda-\mu) \, \mbox{ln}\left(1+\mbox{e}^{-s|\varepsilon_s^{\ast}(\lambda)|}\right) \nonumber \\ 
	&-s& \int_{-Q'}^{-Q} \frac{{\rm d} \mu}{2 \pi} K(\lambda -\mu) \varepsilon_s^{\ast}(\mu)- s \int_{Q}^{Q'} \frac{{\rm d} \mu}{2 \pi} K(\lambda-\mu) \varepsilon_s^{\ast}(\mu) \nonumber \\
	&-& s \int_{-Q}^{Q} \frac{{\rm d} \mu}{2 \pi} K(\lambda-\mu)\left[\varepsilon_s^{\ast}(\mu)-\varepsilon_{\infty}(\mu)\right],
	\label{thirdsteplarges}
\end{eqnarray}
with $Q'$ and $Q$ having the same meaning as in Sec.~\ref{sec:asymptotics_calculations}. The analysis of the integrals appearing on the right hand side is completely analogous to the one carried out in Sec.~\ref{sec:asymptotics_calculations} for Eqs.~\eqref{Qintegralexpansions},\eqref{Qintegralexpansions_2} and \eqref{eq:expansionLowT_2}. In conclusion from Eqs.~\eqref{secondsteplarges},\eqref{thirdsteplarges}, $\delta \varepsilon_s^{\ast}(\lambda)= \varepsilon_s^{\ast}(\lambda)-\varepsilon_\infty(\lambda)$ satisfies the following integral equation
\begin{eqnarray}
	\delta \varepsilon_s^{\ast}(\lambda) = &-&\frac{h_0}{s} -\frac{h_{-1}}{s^2} +\frac{1}{s}\mbox{ln}\left[\frac{\lambda^2}{c^2}\left(\frac{\lambda^2}{c^2} + \frac{1}{4} \right) \right] + \int_{-Q}^{Q}\frac{{\rm d} \mu}{2 \pi} K(\lambda-\mu) \delta \varepsilon_s^{\ast}(\mu) -\frac{\varepsilon'_{\infty}(Q)}{4 \pi}(Q'-Q)^2(K(\lambda-Q)+K(\lambda+Q)) \nonumber \\
	&-& \frac{\pi}{12 \varepsilon'_{\infty}(Q) s^2}[K(\lambda-Q)+K(\lambda+Q)] +O(s^{-3})\,. 
	\label{fourthsteplarges}
\end{eqnarray}
With a reasoning analogous to the one done in Ref.~\cite{BePi2018} one can show that the term containing $(Q'-Q)^2$ is at least of order $O(s^{-2})$. Accordingly,
\begin{equation}
	\delta \varepsilon_s^{\ast}(\lambda) = \frac{U_1(\lambda)}{s} + O(s^{-2}), \label{firstorderUfunction}
\end{equation} 
where the function $U_1(\lambda)$ is the solution of the integral equation
\begin{equation}
	U_1(\lambda) = -h_0 +\mbox{ln}\left[\frac{\lambda^2}{c^2}\left(\frac{\lambda^2}{c^2} +\frac{1}{4}\right)\right] +\int_{-Q}^{Q}\frac{{\rm d} \mu}{2 \pi} K(\lambda -\mu) U_1(\mu)\,. 
\end{equation}
In particular, computing Eq.~\eqref{firstorderUfunction} for $\lambda=Q'$ and expanding it in the difference $Q'-Q$ we get
\begin{equation}
	Q'-Q = -\frac{U_1(Q)}{s \varepsilon'_{\infty}(Q)} + O(s^{-2}) \,.
\end{equation}
Note that this result shows that $(Q'-Q)^2$ is exactly of order $O(s^{-2})$.  As a consequence, in order to determine $\delta \varepsilon_s^{\ast}(\lambda)$ up to the second order in $1/s$ in Eq.~\eqref{fourthsteplarges} we keep the terms containing $Q'-Q$. Exploiting the first-order result reported in Eq.~\eqref{firstorderUfunction}, we finally obtain 
\begin{eqnarray}
	\delta \varepsilon_s^{\ast} (\lambda) &=& \varepsilon_s^{\ast}(\lambda)-\varepsilon_{\infty}(\lambda)= \frac{U_1(\lambda)}{s} + \frac{U_2(\lambda)}{s^2} + O(s^{-3}) \,, \\
	U_2(\lambda) &=& \frac{\left[K(\lambda-Q)+K(\lambda+Q)\right]}{\varepsilon_{\infty}'(Q)}\left(-\frac{U_1^2(Q)}{4 \pi}-\frac{\pi}{12}\right)-h_{-1} + \int_{-Q}^{Q}\frac{{\rm d} \mu}{2 \pi} \, K(\lambda-\mu)U_2(\mu)\,,
\end{eqnarray}
completing the derivation of Eqs.~\eqref{eq:expansionLowT_1}--\eqref{eq:expansionLowT_4} in the main text.

\section{Small-$s$ asymptotics of the scaled cumulant generating function in the Tonks-Girardeau limit}
\label{app:low_s_asymptotic_TG}

In this section we study the small-$s$ asymptotic behavior of the function $h^{\ast}(s, D^2)=h^{\ast}(\tilde{s})$ defined in Eq.~\eqref{eq:def_h_ast}, we name for brevity $\tilde{s}=s D^2$ in this appendix.  

We start by taking the derivative with respect to $\tilde{s}$ of both sides of  Eq.~\eqref{starmultiplierequation}, obtaining
\begin{equation}
	\int_{-\infty}^{\infty} \frac{{\rm d} y}{2 \pi} \frac{y^2 \, \mbox{e}^{2 \tilde{s} y^2 -h^{\ast}}(2 y^2 -\frac{{\rm d} h^{\ast}(\tilde{s})}{{\rm d}\tilde{s}})}{\left(  1+\frac{y^2}{4} \, \mbox{e}^{2\tilde{s} y^2 -h^{\ast}} \right)^2} = 0\,,
\end{equation}
which is equivalent to 
\begin{equation}
	\frac{{\rm d} h^{\ast}(\tilde{s})}{{\rm d} \tilde{s}} \int_{-\infty}^{\infty} \frac{{\rm d} y}{2 \pi} \frac{y^2 \mbox{e}^{2 \tilde{s} y^2 -h^{\ast}(\tilde{s})}}{\left(1+\frac{y^2}{4}\mbox{e}^{2 \tilde{s} y^2 -h^{\ast}} \right)^2} = \int_{-\infty}^{\infty}\frac{{\rm d} y}{2 \pi} \frac{8 y^2 \left(\frac{y^2}{4}\mbox{e}^{2\tilde{s} y^2-h^{\ast}} \right)}{\left(1+\frac{y^2}{4}\mbox{e}^{2\tilde{s}y^2-h^{\ast}}   \right)^2}\,. \label{intermediatestepappendix}
\end{equation}
The right-hand side of this equation can be rewritten as 
\begin{equation}
	\int_{-\infty}^{\infty}\frac{{\rm d} y}{2 \pi} \frac{8 y^2}{1+\frac{y^2}{4}\mbox{e}^{2 \tilde{s}y^2 -h^{\ast}}}-\int_{-\infty}^{\infty} \frac{{\rm d} y}{2 \pi} \frac{8 y^2}{\left(1+\frac{y^2}{4}\mbox{e}^{2\tilde{s} y^2-h^{\ast}} \right)^2} \equiv H_1(\tilde{s}) +H_2(\tilde{s})\,. 
\end{equation}
We analyze the two terms $H_{1,2}(\tilde{s})$ in the  limit $\tilde{s} \rightarrow 0$. For $H_2$, we simply compute
\begin{equation}
	\lim_{\tilde{s}\rightarrow 0} H_2(\tilde{s}) = -\int_{-\infty}^{+\infty} \frac{{\rm d} y}{2 \pi} \frac{8 y^2}{\left(1+\frac{y^2}{4}\right)^2} = -16\,, \label{appendixH2}
\end{equation}
where we have used that $h^{\ast}(\tilde{s}) \rightarrow 0$ as $\tilde{s} \rightarrow 0$, see Eq.~\eqref{eq:def_h_ast}. For $H_1$, instead, we have
\begin{eqnarray}
	H_1(\tilde{s}) &=& \int_{-\infty}^{\infty}\frac{{\rm d} y}{2 \pi} \frac{8 y^2}{1+\frac{y^2}{4} \, \mbox{e}^{2 \tilde{s} y^2}} = 32 \int_{-\infty}^{\infty}\frac{{\rm d} y}{2 \pi} \mbox{e}^{-2 \tilde{s} y^2} \frac{\frac{y^2}{4}}{\frac{y^2}{4} + \mbox{e}^{-2 \tilde{s} y^2}}  \nonumber \\ 
	&=& 32 \int_{-\infty}^{\infty} \frac{{\rm d} y}{2 \pi} \mbox{e}^{-2\tilde{s} y^2} -32 \int_{-\infty}^{\infty}\frac{{\rm d} y}{2 \pi} \frac{\mbox{e}^{-4\tilde{s} y^2}}{\mbox{e}^{-2\tilde{s} y^2}+\frac{y^2}{4}}  \nonumber \\ 
	&=& \frac{16}{\sqrt{2 \pi \tilde{s}}} -32 +O(\tilde{s}^{1/2}) \,. \label{appendixH1}
\end{eqnarray}
Similarly, the integral on the left-hand side of Eq.~\eqref{intermediatestepappendix} can be straightforwardly evaluated for $\tilde{s} \rightarrow 0$. Taking into account Eqs.~\eqref{appendixH2} and \eqref{appendixH1}, Eq.~\eqref{intermediatestepappendix} becomes, for $\tilde{s} \rightarrow 0$,
\begin{equation}
	2 \frac{{\rm d} h^{\ast}(\tilde{s})}{{\rm d} \tilde{s}} = \frac{16}{\sqrt{2 \pi \tilde{s}}} + O\left(\tilde{s}^0\right)\,,
\end{equation}
and therefore, after integration in $\tilde{s}= s D^2$
\begin{equation}
	h^{\ast}(\tilde{s}) = \frac{16}{\sqrt{2 \pi}} \sqrt{\tilde{s}} + O\left(\tilde{s}\right)\,,
\end{equation}
which is Eq.~\eqref{lowshast}. The small-$\tilde{s}$ asymptotic behavior of the scaled cumulant generating function follows from this result. In particular, plugging Eq.~\eqref{lowshast} into Eq.~\eqref{finalresultTG} and performing the change of variable $\lambda = y D$ inside the integral, the latter is rewritten as
\begin{equation}
	\frac{f(\tilde{s})}{D}=1 + \frac{8}{\sqrt{2 \pi}} \sqrt{\tilde{s}} -\frac{\pi^2}{3} \tilde{s} -\frac{1}{2} \int_{-\infty}^{\infty} \frac{\mbox{d} y}{2 \pi} \, \mbox{ln}\left(1+ \frac{4}{y^2} \, \mbox{e}^{-2 \tilde{s} y^2+ 16 \sqrt{\tilde{s}/(2 \pi)}} \right)\,. \label{lowsTonks}
\end{equation}
Taking now the derivative with respect to $\tilde{s}$ of both sides, one has
\begin{eqnarray}
	\frac{{\rm d} (f(\tilde{s})/D)}{{\rm d} \tilde{s}} &=& \frac{4}{\sqrt{2 \pi \tilde{s}}} -\frac{\pi^2}{3}  -\frac{1}{2} \int_{-\infty}^{\infty} \frac{{\rm d} y}{2 \pi} \frac{\frac{4}{y^2}(-2 y^2 +\frac{8}{\sqrt{2 \pi \tilde{s}}})\mbox{e}^{-2\tilde{s}y^2 +16 \sqrt{\tilde{s}/(2 \pi)}}}{1+\frac{4}{y^2} \mbox{e}^{-2\tilde{s}y^2 +16 \sqrt{\tilde{s}/(2 \pi)}}} \nonumber \\ 
	&=& \frac{4}{\sqrt{2 \pi \tilde{s}}} -\frac{\pi^2}{3}  +\frac{1}{2}F'(\tilde{s})\,,
	\label{initialformulalows}
\end{eqnarray}
where $F'(\tilde{s})$ stands for the derivative w.r.t. $\tilde{s}$ of the integral appearing in Eq.~\eqref{lowsTonks}. 
The asymptotics of this integral can be worked out using the same steps as above for the function $h^{\ast}(\tilde{s})$. In particular, defining for convenience $C=16/\sqrt{2 \pi}$, we write
\be
F'(\tilde{s})  = I_1(\tilde{s}) +I_2(\tilde{s})\,,
\ee
where
\begin{eqnarray}
	I_1(\tilde{s}) &=& -4 \int_{-\infty}^{\infty} \frac{{\rm d} y}{2 \pi} \frac{\frac{C}{2 \sqrt{\tilde{s}}}}{y^2 \mbox{e}^{2 \tilde{s} y^2-C\sqrt{\tilde{s}}}+4}\,, \\
	I_2(\tilde{s}) &=& 4 \int_{-\infty}^{\infty} \frac{{\rm d} y}{2 \pi} \frac{2 y^2}{y^2 \mbox{e}^{2\tilde{s} y^2 -C\sqrt{\tilde{s}}}+4}.    
\end{eqnarray}
We analyze $I_1(\tilde{s})$ and $I_2(\tilde{s})$ separately. For $I_1(\tilde{s})$ in the limit $\tilde{s} \rightarrow 0$, we have
\begin{eqnarray}
	I_1(\tilde{s}) &=& -4 \int_{-\infty}^{\infty} \frac{{\rm d} y}{2 \pi} \frac{\frac{C}{2 \sqrt{\tilde{s}}}}{y^2 \mbox{e}^{2 \tilde{s} y^2-C\sqrt{\tilde{s}}}+4} \nonumber \\ 
	&=& -\frac{2 C}{\sqrt{\tilde{s}}}\int_{-\infty}^{\infty}\frac{{\rm d} y}{2 \pi} \frac{1}{y^2 +4} +O(\tilde{s}^0) \nonumber \\
	&=& -\frac{C}{\sqrt{\tilde{s}}} \int_{-\infty}^{\infty} \frac{{\rm d} y}{2 \pi} \frac{1}{1+y^2} +O(\tilde{s}^{0}) \nonumber \\
	&=& -\frac{C}{2 \sqrt{\tilde{s}}}+O(\tilde{s}^{0}) \,, \label{appendix_I1}
\end{eqnarray}
while for $I_2(\tilde{s})$
\begin{eqnarray}
	I_2(\tilde{s}) &=& 8 \int_{-\infty}^{\infty} \frac{{\rm d} y}{2 \pi} \frac{y^2}{y^2 \mbox{e}^{2\tilde{s} y^2 -C\sqrt{\tilde{s}}}+4} \nonumber \\ 
	&=&  8 \int_{-\infty}^{\infty} \frac{{\rm d} y}{2 \pi} \mbox{e}^{-2 \tilde{s} y^2 + C\sqrt{\tilde{s}}}\left(1-\frac{4 \, \mbox{e}^{-2\tilde{s}y^2 +C\sqrt{\tilde{s}}}}{y^2 +4 \, \mbox{e}^{-2\tilde{s}y^2 +C\sqrt{\tilde{s}}}}  \right) \nonumber \\
	& \equiv & J_1(\tilde{s}) +J_2(\tilde{s})\,.
\end{eqnarray}
For $J_{1,2}$ introduced above we have
\begin{eqnarray}
	J_1(\tilde{s}) &=& 8 \mbox{e}^{C \sqrt{\tilde{s}}}\int_{-\infty}^{\infty}\frac{{\rm d} y}{2 \pi} \mbox{e}^{-2\tilde{s} y^2} = \frac{4 }{\sqrt{2 \pi \tilde{s}}} \mbox{e}^{C \sqrt{\tilde{s}}} \nonumber \\
	&=& \frac{4}{\sqrt{2 \pi \tilde{s}}} +O(\tilde{s}^0)\,, \label{appendix_J1}
\end{eqnarray}
while
\begin{eqnarray}
	J_2(\tilde{s}) &=& -8 \int_{-\infty}^{\infty}\frac{{\rm d} y}{2 \pi} \mbox{e}^{-2 \tilde{s}y^2+C \sqrt{\tilde{s}}}\frac{4}{y^2 \mbox{e}^{2\tilde{s}y^2 -C\sqrt{\tilde{s}}} +4} \nonumber \\ 
	&=& -8 \int_{-\infty}^{\infty}\frac{{\rm d} y}{2 \pi} \frac{4}{y^2 +4} +O(\sqrt{\tilde{s}}) = -8 + O(\sqrt{\tilde{s}}).
\end{eqnarray}
Collecting all the terms in Eqs.~\eqref{appendix_I1} and \eqref{appendix_J1}, Eq.~\eqref{initialformulalows} yields
\begin{equation}
	\frac{{\rm d} (f(s)/D)}{{\rm d} \tilde{s}} = \frac{2}{\sqrt{2 \pi \tilde{s}}}+O(\tilde{s}^0)\,.
\end{equation}
After integration, we finally obtain
\begin{equation}
	\frac{f(s)}{D} = 2 \sqrt{\frac{2}{\pi}} \, \, \tilde{s}^{1/2} + O(\tilde{s})\,,
\end{equation}
i.e., Eq.~\eqref{lowsasym}.

\section{Large-$s$ asymptotics of the scaled cumulant generating function in the Tonks-Girardeau limit}
\label{app:large_s_asymptotic_TG}

In this appendix we provide details of the calculation leading to the large-$s$ expansion of the scaled cumulant generating function $f(s)$ reported in Eq.~\eqref{eq:large_s_asympt_TG}. As well as in Appendix.~\ref{app:low_s_asymptotic_TG} we denote for brevity $\tilde{s}=s D^2$.

We start from Eq.~\eqref{finalresultTG} with $h^{\ast}(\tilde{s})$ given by Eq.~\eqref{largesmultiplierasymptotic} and $h^{\prime}=2 \pi^2$. One rewrites it as
\be
\frac{f(\tilde{s})}{D} = 1 + \frac{2 \pi^2}{3} \tilde{s} + \frac{h^{\prime}_0}{2} + \frac{f_{>}(\tilde{s})}{D} + \frac{f_{<}(\tilde{s})}{D}\,,
\ee
where
\bea
\frac{f_{>}(\tilde{s})}{D} &=& -\int_{\pi}^{\infty} \frac{{\rm d} y}{2 \pi} \, \mbox{ln} \left( 1 + \frac{4}{y^2}\mbox{e}^{-\tilde{s}(2 y^2-h^{\prime})} e^{h^{\prime}_0} \right)\,, \label{fgreater} \\ 
\frac{f_{<}(\tilde{s})}{D} &=& -\int_{0}^{\pi} \frac{{\rm d} y}{2 \pi} \, \mbox{ln} \left( 1 + \frac{4}{y^2}\mbox{e}^{-\tilde{s}(2 y^2-h^{\prime})} e^{h^{\prime}_0} \right)\,, \label{flesser}.
\eea
where we performed the change of variable $\lambda = D y$ inside the integral in Eq.~\eqref{finalresultTG}. The value $\lambda=Q= D \pi$ ($y=\pi$) is the TG limit of the analogous symbol introduced in Sec.~\ref{sec:asymptotics_calculations} and then in Appendix \ref{app:large_s_asymptotic}.
First, we rewrite $f_{<}(\tilde{s})/D$ as
\begin{eqnarray}
	\frac{f_{<}(\tilde{s})}{D} &=& -\int_{0}^{\pi} \frac{{\rm d} y}{2 \pi} \, \mbox{ln} \left(1+\frac{4}{y^2}\mbox{e}^{-\tilde{s}(2 y^2-h^{\prime})} e^{h^{\prime}_0} \right)  \nonumber \\ 
	&=& 2 f_0 -1 - \frac{2 \pi^2}{3} \tilde{s} - \frac{h^{\prime}_0}{2} -\int_{0}^{\pi} \frac{{\rm d} y}{2 \pi} \, \mbox{ln} \left( 1+ \frac{y^2}{4}\mbox{e}^{\tilde{s}(2 y^2-h^{\prime})} e^{-h^{\prime}_0} \right)\,,
\end{eqnarray}
with $f_0$ given in Eq.~\eqref{largesasym0}. Expanding the logarithm in a power series, one has:
\begin{eqnarray}
	-\int_{0}^{\pi} \frac{{\rm d} y}{2 \pi} \, \mbox{ln} \left( 1+ \frac{y^2}{4}\mbox{e}^{\tilde{s}(2 y^2-h^{\prime})}\mbox{e}^{-h^{\prime}_0} \right) &=& -\sum_{n=1}^{+ \infty}\frac{(-1)^{n+1}\mbox{e}^{-n\tilde{s} h^{\prime}} \mbox{e}^{-n h^{\prime}_0} }{2 \pi  n \, 4^{n}} \int_{0}^{\pi} {\rm d} y \, y^{2n} \mbox{e}^{n \tilde{s} 2 y^2}\,,
	\label{integrallarges}
\end{eqnarray}
where the last integral, after integration by parts, can be estimated as
\begin{eqnarray}
	\int_{0}^{\pi} {\rm d} y \, y^{2n} \mbox{e}^{n \tilde{s} 2 y^2} &=& \frac{1}{4 n \tilde{s} (2 n \tilde{s})^{n-1/2} } \int_{0}^{2 n \tilde{s} \pi^2} {\rm d}z \, z^{n-1/2}\mbox{e}^z \nonumber \\
	&=& \frac{\pi^{2n-1}}{4 n \tilde{s}}\mbox{e}^{2 n \tilde{s} \pi^2} + O\left( \frac{\mbox{e}^{2 n \tilde{s} \pi^2}}{\tilde{s}^2} \right)\,.
	\label{eq:estimate}
\end{eqnarray}
Plugging Eq.~\eqref{eq:estimate} into Eq.~\eqref{integrallarges}, we get
\begin{eqnarray}
	-\int_{0}^{\pi} \frac{{\rm d} y}{2 \pi} \, \mbox{ln} \left( 1+ \frac{y^2}{4}\mbox{e}^{\tilde{s}(2 y^2-h^{\prime})} \mbox{e}^{-h^{\prime}_0} \right) &=& \frac{1}{8 \tilde{s} \pi^2}\sum_{n=1}^{+ \infty}\frac{(-1)^{n} \pi^{2n} \mbox{e}^{-n h^{\prime}_0}}{4^{n} n^2} +O(\tilde{s}^{-2}) \nonumber \\
	&=& \frac{1}{8 \tilde{s} \pi^2} \mbox{Li}_2\left(-\frac{\pi^2}{4} \mbox{e}^{-h^{\prime}_0} \right) +O(\tilde{s}^{-2}). \label{firstpart}
\end{eqnarray}
The evaluation of $f_{>}(\tilde{s})/D$ proceeds along the same lines: in particular, after expanding the logarithm, we can write
\begin{eqnarray}
	\frac{f_{>}(\tilde{s})}{D} &=& -\sum_{n=1}^{+\infty} \frac{(-1)^{n+1} 4^{n}\mbox{e}^{ n \tilde{s} h^{\prime}} \mbox{e}^{n h^{\prime}_0}}{2 \pi n}\int_{\pi}^{+\infty} {\rm d}y \, \frac{\mbox{e}^{-\tilde{s} n 2y^2}}{y^{2n}} \nonumber \\
	&=& \frac{1}{8 \tilde{s} \pi^2} \sum_{n=1}^{+\infty}  \frac{(-4)^n \mbox{e}^{n h^{\prime}_0} }{\pi^{2n} n^2} +O(\tilde{s}^{-2}) = \frac{1}{8 \tilde{s} \pi^2} \mbox{Li}_2\left(-\frac{4}{\pi^2} \mbox{e}^{h^{\prime}_0} \right) + O(\tilde{s}^{-2})\,. \label{secondpart}
\end{eqnarray}
Summing now the results in Eqs.~\eqref{firstpart}, \eqref{secondpart}, and using the identity (see, e.g., Ref.~\cite{NIST})
\begin{equation}
	\mbox{Li}_2\left(z\right) + \mbox{Li}_2\left(\frac{1}{z}\right) = -\frac{\pi^2}{6} -\frac{1}{2} \mbox{ln}^2\left(-z \right)  \quad z \in \mathbb{C}  \, \backslash \, (1,+\infty)\,,
\end{equation}
we finally arrive at Eq.~\eqref{eq:large_s_asympt_TG}.

\section{Algebraic behavior of $p(w)$ for large $w$:  arbitrary particle number}
\label{app:power_law}

In this appendix we show how to extend by induction the computation presented in Sec.~\ref{sec:power_law_tail} for $N=3$ to an arbitrary number $N$ of particles. In order to simplify the discussion we will first present the explicit example $M=2$ ($N=5$), and then treat the general case. As a result of the analysis of this appendix, we conclude that Eq.~\eqref{onerapiditydistribution_intensive} holds for arbitrary values of $N$.

\begin{figure}
	\centering
	\includegraphics[width=0.5\columnwidth]{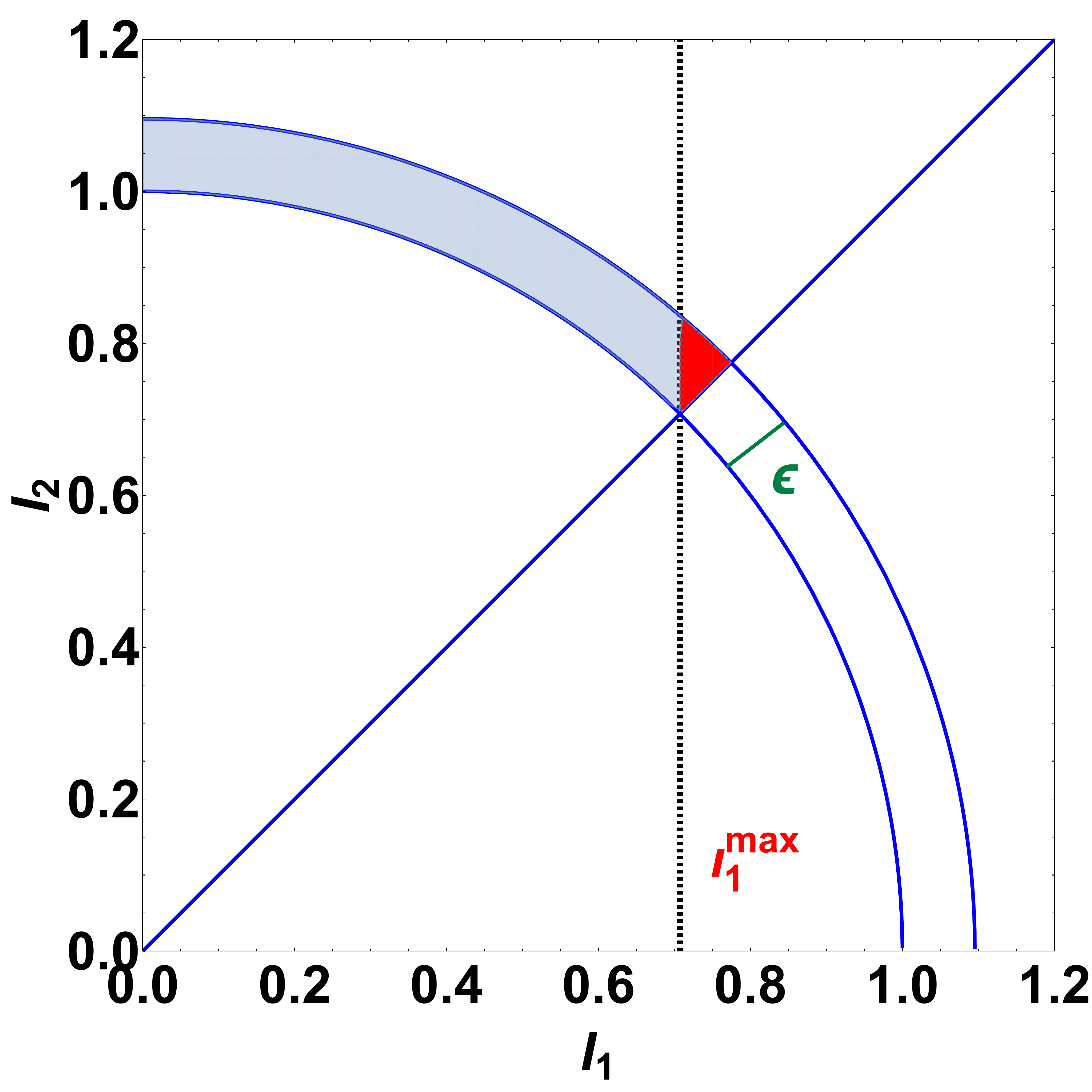}	
	\caption{Pictorial representation of the summation domain $\mathcal{D}$ of Eq.~\eqref{summationdomainformula} in blue and red shaded areas. The latter one is of order $\epsilon^2$, with $\epsilon$ the thickness of the shell, and can therefore be neglected to order $\epsilon$. The intercept of the vertical black dashed line with the horizontal $I_1$ axis is given by  $I_1^{\rm max}= L \sqrt{(W-\epsilon)}/(4 \pi)$.}
	\label{fig:summationdomain}
\end{figure}

In the case $M=2$, Eq.~\eqref{workshelldefinition} reads
\begin{eqnarray}
	\epsilon P(W)&=& \sum_{W' \in (W-\epsilon, W+\epsilon)} |\langle I_1,I_2 | {\rm BEC} \rangle|^2 \nonumber \\ 
	&=& \sum_{I_1,I_2 \in \mathcal{D}} |\langle I_1,I_2 |{\rm BEC} \rangle|^2, \label{eq:app_starting_two_rapidities}
\end{eqnarray}
where the domain $\mathcal{D}$ of the double sum is determined, via Eq.~\eqref{eq:small_density_eigenvalues}, by the boundaries of the energy shell in Eq.~\eqref{eq:app_starting_two_rapidities} and the fact the two quantum numbers $I_1$ and $I_2$ have to be different. In particular \\[0.3mm]
\begin{equation}
	I_1,I_2 \in \mathcal{D}\Leftrightarrow \left\{
	\begin{array}{ll}
		I_1^2 +I_2^2 \in \frac{L^2}{8 \pi^2}(W-\epsilon,W+\epsilon) \\
		0< I_1<I_2,
	\end{array}
	\right.
	\label{summationdomainformula}
\end{equation}
\\[0.3mm]
corresponding to the region highlighted in blue in the $I_1-I_2$ plane shown in Fig.~\ref{fig:summationdomain}. For convenience we will neglect from the start the contribution of the domain $\mathcal{D}$ where $I_1>I_1^{\rm max} = L \sqrt{(W-\epsilon)}/(4 \pi) $, depicted in red in Fig.~\ref{fig:summationdomain}, which provides a contribution $O(\epsilon^2)$. Then, from Eqs.~\eqref{eq:app_starting_two_rapidities},\eqref{eq:smalldensityrapidities} and \eqref{smalldensityoverlaps} one has 
\begin{equation}
	\epsilon P(W) = \frac{5! c^4 L^4}{(16 \pi^2)^2} \sum_{I_1=1}^{I_1^{\rm max}} \frac{1}{I_1^2(\pi^2 I_1^2 +\frac{c^2 L^2}{16})} \sum_{I_2=I_2^{\rm min}(I_1)}^{I_2^{\rm max}(I_1)} \frac{1}{I_2^2(\pi^2 I_2^2 +\frac{c^2 L^2}{16})}\,,
	\label{tworapiditiesstartingpoint}
\end{equation}
where $I_2=I_2^{\rm min}(I_1)$ and $I_2^{\rm max}(I_1)$ are obtained from the on-shell condition in Eq.~\eqref{summationdomainformula}, in analogy with the case $M=1$, and read
\begin{eqnarray}
	I_2^{\rm min}(I_1) &=& \sqrt{\frac{L^2}{8 \pi^2} {W-\epsilon}-I_1^2}, \nonumber \\
	I_2^{\rm max}(I_1) &=& \sqrt{\frac{L^2}{8 \pi^2} {W+\epsilon}-I_1^2}, \nonumber \\
	\delta I(I_1) &\equiv& I_2^{\rm max}(I_1)-I_2^{\rm min}(I_1) =  \frac{L^2 \epsilon}{8 \pi^2 \sqrt{\frac{L^2 W}{8 \pi^2}-I_1^2}} +O\left( \epsilon^2 \right)\,.
	\label{shellIvalues2}
\end{eqnarray}
The sum $\mathcal{S}(I_1)$ over $I_2$ in Eq.~\eqref{tworapiditiesstartingpoint} can be bounded using the same argument as the one presented in Sec.~\ref{sec:power_law_tail} for the case $M=1$ to get Eqs.~\eqref{pminbounds} and \eqref{pmaxbounds}; in particular at leading order in $\epsilon$
\begin{eqnarray}
	\mathcal{S}(I_1) & \equiv & \sum_{I_2=I_2^{\rm min}(I_1)}^{I_2^{\rm max}(I_1)} \frac{1}{I_2^2(\pi^2 I_2^2 +\frac{c^2 L^2}{16})} \nonumber \\
	&=&  \sum_{j=0}^{\delta I(I_1)} \frac{1}{[j+I_2^{\rm min}(I_1)]^2[\pi^2 (j+I_2^{\rm min}(I_1))^2 +\frac{c^2 L^2}{16}]} \nonumber \\
	&=& \frac{L^2 \epsilon}{8 \pi^2 \sqrt{L^2 W /(8 \pi^2) -I_1^2}}\frac{1}{L^2 W /(8 \pi^2) -I_1^2} \frac{1}{\pi^2(L^2 W /(8 \pi^2) -I_1^2)+\frac{c^2 L^2}{16}} + O(\epsilon^2)\,.
	\label{firstboundtworapidities}
\end{eqnarray}
In terms of $\mathcal{S}(I_1)$ in Eq.~\eqref{firstboundtworapidities}, $\epsilon P(w)$ in Eq.~\eqref{tworapiditiesstartingpoint} can be written as
\begin{equation}
	\epsilon P(W) = \frac{5! c^4 L^4}{(16 \pi^2)^2} \sum_{I_1=1}^{I_1^{\rm max}} \frac{1}{I_1^2(\pi^2 I_1^2 +\frac{c^2 L^2}{16})} \mathcal{S}(I_1)\,,
\end{equation}
where $\mathcal{S}(I_1)$ can be bounded as
\begin{eqnarray}
	\mathcal{S}(I_1=1) &<& \mathcal{S}(I_1) < \mathcal{S}(I_1^{\rm max}), \nonumber \\ 
	\mathcal{S}(I_1=1) & \propto & \mathcal{S}(I_1^{\rm max}) \propto \frac{L^2 \epsilon}{\sqrt{L^2 W} L^2 W (\frac{L^2 W}{8 \pi^2}+\frac{c^2 L^2}{16})} + O(\epsilon^2)\,, \label{Ssumbound} 
\end{eqnarray}
where the symbol $\propto$ henceforth indicates that we are neglecting numerical prefactors. From Eq.~\eqref{Ssumbound} it follows that
\begin{equation}
	\epsilon P(W) \propto c^4 L^4 \frac{L^2 \epsilon}{\sqrt{L^2 W} L^2 W (\frac{L^2 W}{8 \pi^2}+\frac{c^2 L^2}{16})} \sum_{I_1=1}^{L \sqrt{W}/4 \pi} \frac{1}{I_1^2(\pi^2 I_1^2 +\frac{c^2 L^2}{16})} +O(\epsilon^2), \label{almostfinishedpowerlawtail}
\end{equation}
and, equivalently, for the probability density function $p(w)=LP(w L)$ of the intensive work $w$
\begin{equation}
	\epsilon p(w) \propto c^4 L^5 \frac{L^2 \epsilon}{\sqrt{L^3 w} L^3 w (\frac{L^3 w}{8 \pi^2}+\frac{c^2 L^2}{16})} \sum_{I_1=1}^{L \sqrt{w L}/4 \pi} \frac{1}{I_1^2(\pi^2 I_1^2 +\frac{c^2 L^2}{16})} +O(\epsilon^2). \label{almostfinishedpowerlawtail2}
\end{equation}
In order to get the leading behaviour of $p(w)$ as a function of $L$, as a final step, we need to estimate the asymptotic of the sum over $I_1$ in Eq.~\eqref{almostfinishedpowerlawtail2} for $L$ large. We have
\begin{equation}
	\sum_{I_1=1}^{L \sqrt{w L}/4 \pi} \frac{1}{I_1^2(\pi^2 I_1^2 +\frac{c^2 L^2}{16})} = \sum_{I_1=1}^{\infty} \frac{1}{I_1^2(\pi^2 I_1^2 +\frac{c^2 L^2}{16})} -\sum_{I_1=L \sqrt{w L}/4 \pi}^{\infty} \frac{1}{I_1^2(\pi^2 I_1^2 +\frac{c^2 L^2}{16})}, \label{lastsuminduction1}
\end{equation}
with 
\begin{eqnarray}
	\sum_{I_1= L \sqrt{w L}/4 \pi}^{\infty} \frac{1}{I_1^2(\pi^2 I_1^2 +\frac{c^2 L^2}{16})} \leq  \frac{1}{L^3 w/(16 \pi^2)} \sum_{I_1= L \sqrt{w L}/4 \pi}^{\infty} \frac{1}{(\pi^2 I_1^2 +\frac{c^2 L^2}{16})} &\leq & \frac{16 \pi^2}{L^3 w} \sum_{I_1=1}^{\infty} \frac{1}{(\pi^2 I_1^2 +\frac{c^2 L^2}{16})} \nonumber \\ 
	&=& \frac{32 \pi^2}{c \, w L^4} +O(L^{-5}) \quad \mbox{as} \, \, L \rightarrow \infty \label{lastsuminduction2}
\end{eqnarray}
and
\begin{eqnarray}
	\sum_{I_1=1}^{\infty} \frac{1}{I_1^2(\pi^2 I_1^2 +\frac{c^2 L^2}{16})} = 8 \pi^2 \frac{48 +c^2 L^2 -12 \, c L \, \mbox{coth}\left(\frac{c L}{4}\right)}{3 c^4 L^4} = \frac{8 \pi^2}{3 c^2 L^2}+ O(L^{-3})  \quad \mbox{as} \, \, L \rightarrow \infty. \label{lastsuminduction3}
\end{eqnarray}
Plugging this into Eq.~\eqref{almostfinishedpowerlawtail2} we arrive at the final result 
\begin{eqnarray}
	p(w) &\propto & c^4 L^5 \frac{L^2}{\sqrt{L^3 w} L^3 w (\frac{L^3 w}{8 \pi^2}+\frac{c^2 L^2}{16})}\left[\frac{8 \pi^2}{3 c^2 L^2}+O(L^{-3}) \right]. \nonumber \\
	&\propto & \frac{c^2}{L^{5/2}} w^{-5/2} + O(L^{-7/2}) \quad \mbox{as} \, \, L \rightarrow \infty.
	\label{tworapiditiespowerlawtail} 
\end{eqnarray}

At this point it should be clear how to generalize the result in Eq.~\eqref{tworapiditiespowerlawtail} by induction to the general case of $N=2M+1$ particles. Indeed, consider the expression for $P_M(W)$ 
\begin{eqnarray}
	\epsilon P_M(W) &=& \sum_{I_1,I_2,...I_M \in \mathcal{D}} |\langle I_1,I_2,...I_M | {\rm BEC} \rangle|^2 \nonumber \\
	&=& \frac{N c^2 L^2}{16 \pi^2} \sum_{I_1=1}^{I_1^{\rm max}} \frac{1}{I_1^2(\pi^2 I_1^2 +\frac{c^2 L^2}{16})} \left(\sum_{I_2,I_3,...I_M \in \mathcal{D'}(W')} |\langle I_2,I_3,...I_M | {\rm BEC} \rangle|^2\right)\,.
	\label{inductionMrapiditiesprobability}
\end{eqnarray}
Here the sum in the first line is over the $M$-dimensional shell $\mathcal{D}$ defined by \\[0.3mm]
\begin{equation}
	I_1\ldots I_M\in  \mathcal{D}\Leftrightarrow \left\{
	\begin{array}{ll}
		I_1^2 + I_2^2 + ...I_M^2 \in \frac{L^2}{8 \pi^2}(W-\epsilon,W+\epsilon), \\
		0<I_1<I_2...<I_M,
	\end{array}
	\right. \label{inductionsummationdomainformula}
\end{equation}
\\[0.3mm]
while $\mathcal{D'}$, appearing in the second line, is the same as $\mathcal{D}$ with $W \rightarrow W'= W -8 \pi^2 I_1^2/L^2$ and $I_1^{\rm max}=L \sqrt{(W-\epsilon)}/(\sqrt{8 M} \pi)$. Then, we note that Eq.~\eqref{inductionMrapiditiesprobability} can be written in terms of $P_{M-1}$ as 
\begin{equation}
	\epsilon P_M(W) = \frac{N c^2 L^2}{16 \pi^2} \sum_{I_1=1}^{I_1^{\rm max}} \frac{1}{I_1^2(\pi^2 I_1^2 +\frac{c^2 L^2}{16})} \epsilon P_{M-1}(W').
\end{equation} 
Exploiting now the induction hypothesis, we have
\begin{equation}
	\epsilon P_{M-1}(W') \propto c^2 L^2 \frac{L^2 \epsilon}{\sqrt{L^2 W} L^2 W (\frac{L^2 W}{8 \pi^2}+\frac{c^2 L^2}{16})} +O(\epsilon^2)
\end{equation}
where again we neglect numerical prefactors depending on $N$. Then using Eqs.~\eqref{lastsuminduction1}, \eqref{lastsuminduction2} and \eqref{lastsuminduction3} it follows that $p_M(w)$ displays the large-$L$ behavior in Eq.~\eqref{tworapiditiespowerlawtail}, concluding the derivation. 


\twocolumngrid


\begin{thebibliography}{99}
	
	
	\bibitem{CaEM16} 
	P. Calabrese, F. H. L. Essler, and G. Mussardo, 
	\href{http://dx.doi.org/10.1088/1742-5468/2016/06/064001}{J. Stat. Mech. (2016) 064001}.
	
	
	\bibitem{HLSI08} 
	S. Hofferberth, I. Lesanovsky, T. Schumm, A. Imambekov, V. Gritsev, E. Demler, and J. Schmiedmayer, 
	\href{http://dx.doi.org/10.1038/nphys941}{Nature Phys. {\bf 4}, 489 (2008)}.
	
	\bibitem{AJKB10} 
	J. Armijo, T. Jacqmin, K. V. Kheruntsyan, and I. Bouchoule, 
	\href{http://dx.doi.org/10.1103/PhysRevLett.105.230402}{Phys. Rev. Lett. {\bf 105}, 230402 (2010)}.
	
	\bibitem{JABK11} 
	T. Jacqmin, J. Armijo, T. Berrada, K. V. Kheruntsyan, and I. Bouchoule, 
	\href{http://dx.doi.org/10.1103/PhysRevLett.106.230405}{Phys. Rev. Lett. {\bf 106}, 230405 (2011)}.
	
	\bibitem{KPIS10} 
	T. Kitagawa, S. Pielawa, A. Imambekov, J. Schmiedmayer, V. Gritsev, and E. Demler, 
	\href{http://dx.doi.org/10.1103/PhysRevLett.104.255302}{Phys. Rev. Lett. {\bf 104}, 255302 (2010)}.
	
	\bibitem{KISD11} 
	T. Kitagawa, A. Imambekov, J. Schmiedmayer, and E. Demler, 
	\href{http://dx.doi.org/10.1088/1367-2630/13/7/073018}{New J. Phys. {\bf 13}, 073018 (2011)}.
	
	
	\bibitem{ChDe07} 
	R. W. Cherng and E. Demler, 
	\href{http://dx.doi.org/10.1088/1367-2630/9/1/007}{New J. Phys. {\bf 9}, 7 (2007)}.
	
	\bibitem{LaFe08} 
	A. Lamacraft and P. Fendley, 
	\href{http://dx.doi.org/10.1103/PhysRevLett.100.165706}{Phys. Rev. Lett. {\bf 100}, 165706 (2008)}.
	
	\bibitem{ia-13}
	D. A. Ivanov and A. G. Abanov,
	\href{http://dx.doi.org/10.1103/PhysRevE.87.022114}{Phys. Rev. E {\bf 87}, 022114 (2013)}.
	
	\bibitem{sk-13}
	Y. Shi and I. Klich, 
	\href{http://dx.doi.org/10.1088/1742-5468/2013/05/P05001}{J. Stat. Mech. (2013) P05001}.
	
	\bibitem{e-13}
	V. Eisler, 
	\href{https://doi.org/10.1103/PhysRevLett.111.080402}{Phys. Rev. Lett. {\bf 111}, 080402 (2013)}.
	
	\bibitem{k-14}
	I. Klich, 
	\href{http://dx.doi.org/10.1088/1742-5468/2014/11/P11006}{J. Stat. Mech. (2014) P11006}.
	
	\bibitem{mcsc-15}
	M. Moreno-Cardoner, J. F. Sherson and G. De Chiara, 
	\href{http://dx.doi.org/10.1088/1367-2630/18/10/103015}{New J. Phys. {\bf 18}, 103015 (2016)}.
	
	\bibitem{sp-17}
	J.-M. St\'ephan and F. Pollmann, 
	\href{http://dx.doi.org/10.1103/PhysRevB.95.035119}{Phys. Rev. B {\bf 95}, 035119 (2017)}.
	
	\bibitem{CoEG17} 
	M. Collura, F. H. L. Essler, and S. Groha, 
	\href{http://dx.doi.org/10.1088/1751-8121/aa87dd}{J. Phys. A {\bf 50}, 414002 (2017)}.
	
	\bibitem{nr-17}
	K. Najafi and M. A. Rajabpour, 
	\href{http://dx.doi.org/10.1103/PhysRevB.96.235109}{Phys. Rev. B {\bf 96}, 235109 (2017)}.
	
	\bibitem{hb-17}
	S. Humeniuk and H. P. B\"uchler, 
	\href{http://dx.doi.org/10.1103/PhysRevLett.119.236401}{Phys. Rev. Lett. {\bf 119}, 236401 (2017)}.
	
	\bibitem{EiRW03} 
	V. Eisler, Z. Rácz, and F. van Wijland, 
	\href{http://dx.doi.org/10.1103/PhysRevE.67.056129}{Phys. Rev. E {\bf 67}, 056129 (2003)}.
	
	\bibitem{gadp-06}
	V. Gritsev, E. Altman, E. Demler and A. Polkovnikov, 
	\href{http://dx.doi.org/10.1038/nphys410}{Nature Phys. {\bf 2}, 705 (2006)}.
	
	\bibitem{er-13}
	V. Eisler and Z. R\'acz, 
	\href{https://doi.org/10.1103/PhysRevLett.110.060602}{Phys. Rev. Lett. {\bf 110}, 060602 (2013)}.
	
	\bibitem{lddz-15}
	I. Lovas, B. Dora, E. Demler, and G. Zarand, \href{http://dx.doi.org/10.1103/PhysRevA.95.053621}{Phys. Rev. A {\bf 95}, 053621 (2017)}. 
	
	\bibitem{bpc-18}
	A. Bastianello, L. Piroli, and P. Calabrese, 
	\href{http://dx.doi.org/10.1103/PhysRevLett.120.190601}{Phys. Rev. Lett. {\bf 120}, 190601 (2018)}.
	
	\bibitem{BaPi18}
	A. Bastianello and L. Piroli, 
	\href{http://dx.doi.org/10.1088/1742-5468/aaeb48}{J. Stat. Mech. (2018) 113104}.
	
	\bibitem{GrEC18} 
	S. Groha, F. Essler, and P. Calabrese, 
	\href{http://dx.doi.org/10.21468/SciPostPhys.4.6.043}{SciPost Phys. {\bf 4}, 043 (2018)}.
	
	\bibitem{CoEs19} 
	M. Collura and F. H. L. Essler, 
	\href{http://arxiv.org/abs/1901.04402}{arXiv:1901.04402 (2019)}.
	
	
	\bibitem{Jarz97} 
	C. Jarzynski, 
	\href{http://dx.doi.org/10.1103/PhysRevLett.78.2690}{Phys. Rev. Lett. {\bf 78}, 2690 (1997)}.
	
	\bibitem{TalLutz07}
	P. Talkner, E. Lutz, and P. Hänggi, \href{https://journals.aps.org/pre/abstract/10.1103/PhysRevE.75.050102}{Phys. Rev. E {\bf 75}, 050102(R) (2007)}.  
	
	\bibitem{Silv08} 
	A. Silva, 
	\href{http://dx.doi.org/10.1103/PhysRevLett.101.120603}{Phys. Rev. Lett. {\bf 101}, 120603 (2008)}.
	
	\bibitem{GooldPlastina18}
	J. Goold, F. Plastina, A. Gambassi, A. Silva, 
	in: F. Binder, L. Correa, C. Gogolin, J. Anders, and G. Adesso (eds), Thermodynamics in the Quantum Regime. Fundamental Theories of Physics, \href{https://doi.org/10.1007/978-3-319-99046-0_13}{ {\bf 195}, 317 (Springer, Cham, 2019)}.
	
	
	
	\bibitem{cc-06}
	P. Calabrese and J. Cardy, 
	\href{http://dx.doi.org/10.1103/PhysRevLett.96.136801}{Phys. Rev. Lett. {\bf 96}, 136801 (2006)};\\
	P. Calabrese and J. Cardy, 
	\href{http://dx.doi.org/10.1088/1742-5468/2007/06/P06008}{J. Stat. Mech. (2007) P06008}.
	
	\bibitem{PolkSeng11}
	A. Polkovnikov, K. Sengupta, A. Silva, and M. Vengalattore, \href{https://journals.aps.org/rmp/abstract/10.1103/RevModPhys.83.863}{Rev. Mod. Phys. {\bf 83}, 863 (2011)}.
	
	
	\bibitem{PaSi09}
	F. N. C. Paraan and A. Silva, \href{https://journals.aps.org/pre/abstract/10.1103/PhysRevE.80.061130}{Phys. Rev. E {\bf 80}, 061130 (2009)}. 
	
	\bibitem{SmSi13} 
	P. Smacchia and A. Silva, 
	\href{http://dx.doi.org/10.1103/PhysRevE.88.042109}{Phys. Rev. E {\bf 88}, 042109 (2013)}.
	
	\bibitem{MaSi14}
	J. Marino and A. Silva, \href{https://journals.aps.org/prb/abstract/10.1103/PhysRevB.89.024303}{Phys. Rev. B {\bf 89}, 024303 (2014)}. 
	
	\bibitem{Rotondo18}
	P. Rotondo, J. Min{\'a}{\v{r}}, J. P. Garrahan, I. Lesanovsky, and M. Marcuzzi, \href{https://journals.aps.org/prb/abstract/10.1103/PhysRevB.98.184303}{Phys. Rev. B { \bf 98}, 184303 (2018)}. 
	
	\bibitem{PaSo14} 
	T. P\'almai and S. Sotiriadis, 
	\href{http://dx.doi.org/10.1103/PhysRevE.90.052102}{Phys. Rev. E {\bf 90}, 052102 (2014)}.
	
	\bibitem{Palm15} 
	T. Palmai, 
	\href{http://dx.doi.org/10.1103/PhysRevB.92.235433}{Phys. Rev. B {\bf 92}, 235433 (2015)}.
	
	\bibitem{BDKP11} 
	G. Bunin, L. D’Alessio, Y. Kafri, and A. Polkovnikov, 
	\href{http://dx.doi.org/10.1038/nphys2057}{Nature Phys. {\bf 7}, 913 (2011)}.
	
	\bibitem{GaSi11} 
	A. Gambassi and A. Silva, 
	\href{https://arxiv.org/abs/1106.2671}{arXiv:1106.2671 (2011)}.
	
	\bibitem{GaSi12} 
	A. Gambassi and A. Silva, 
	\href{http://dx.doi.org/10.1103/PhysRevLett.109.250602}{Phys. Rev. Lett. {\bf 109}, 250602 (2012)}.
	
	\bibitem{SoGS13} 
	S. Sotiriadis, A. Gambassi, and A. Silva, 
	\href{http://dx.doi.org/10.1103/PhysRevE.87.052129}{Phys. Rev. E {\bf 87}, 052129 (2013)}.
	
	
	\bibitem{HePK13} 
	M. Heyl, A. Polkovnikov, and S. Kehrein, 
	\href{http://dx.doi.org/10.1103/PhysRevLett.110.135704}{Phys. Rev. Lett. {\bf 110}, 135704 (2013)}.
	
	\bibitem{Heyl14} 
	M. Heyl, 
	\href{http://dx.doi.org/10.1103/PhysRevLett.113.205701}{Phys. Rev. Lett. {\bf 113}, 205701 (2014)}.
	
	\bibitem{Heyl18} 
	M. Heyl, 
	\href{http://dx.doi.org/10.1088/1361-6633/aaaf9a}{Rep. Prog. Phys. {\bf 81}, 054001 (2018)}.
	
	\bibitem{AbKe16} 
	N. O. Abeling and S. Kehrein, 
	\href{http://dx.doi.org/10.1103/PhysRevB.93.104302}{Phys. Rev. B {\bf 93}, 104302 (2016)}.
	
	\bibitem{RyAn18} 
	C. Rylands and N. Andrei, 
	\href{http://dx.doi.org/10.1103/PhysRevB.99.085133}{Phys. Rev. B {\bf 99}, 085133 (2019)}.
	
	
	
	\bibitem{Touc09} 
	H. Touchette, 
	\href{http://dx.doi.org/10.1016/j.physrep.2009.05.002}{Phys. Rep. {\bf 478}, 1 (2009)}.
	
	\bibitem{Huang}
	K. Huang, Introduction to statistical physics (Chapman and Hall/CRC, 2009).
	
	\bibitem{CaEs13} 
	J.-S. Caux and F. H. L. Essler, 
	\href{http://dx.doi.org/10.1103/PhysRevLett.110.257203}{Phys. Rev. Lett. {\bf 110}, 257203 (2013)}.
	
	\bibitem{DWBC14} 
	J. De Nardis, B. Wouters, M. Brockmann, and J.-S. Caux, 
	\href{http://dx.doi.org/10.1103/PhysRevA.89.033601}{Phys. Rev. A {\bf 89}, 033601 (2014)}.
	
	\bibitem{Caux16} 
	J.-S. Caux, 
	\href{http://dx.doi.org/10.1088/1742-5468/2016/06/064006}{J. Stat. Mech. (2016) 064006}.
	
	\bibitem{LiLi63} 
	E. Lieb and W. Liniger, 
	\href{http://dx.doi.org/10.1103/PhysRev.130.1605}{Phys. Rev. {\bf 130}, 1605 (1963)}.
	
	\bibitem{Lieb63} 
	E. Lieb, 
	\href{http://dx.doi.org/10.1103/PhysRev.130.1616}{Phys. Rev. {\bf 130}, 1616 (1963)}.
	
	
	
	\bibitem{KiWW04} 
	T. Kinoshita, T. Wenger, and D. S. Weiss, 
	\href{http://dx.doi.org/10.1126/science.1100700}{Science {\bf 305}, 1125 (2004)}.
	
	\bibitem{KiWW06} 
	T. Kinoshita, T. Wenger, and D. S. Weiss, 
	\href{http://dx.doi.org/10.1038/nature04693}{Nature {\bf 440}, 900 (2006)}.
	
	\bibitem{AEWK08} 
	A. H. van Amerongen, J. J. P. van Es, P. Wicke, K. V. Kheruntsyan, and N. J. van Druten, 
	\href{http://dx.doi.org/10.1103/PhysRevLett.100.090402}{Phys. Rev. Lett. {\bf 100}, 090402 (2008)}.
	
	\bibitem{HGMD09} 
	E. Haller, M. Gustavsson, M. J. Mark, J. G. Danzl, R. Hart, G. Pupillo, and H.-C. N\"agerl, 
	\href{http://dx.doi.org/10.1126/science.1175850}{Science {\bf 325}, 1224 (2009)}.
	
	\bibitem{EWAR10} 
	J. J. P. van Es, P. Wicke, A. H. van Amerongen, C. R\'etif, S. Whitlock, and N. J. van Druten, 
	\href{http://dx.doi.org/10.1088/0953-4075/43/15/155002}{J. Phys. B {\bf 43}, 155002 (2010)}.
	
	\bibitem{KHML10} 
	P. Kr\"uger, S. Hofferberth, I. E. Mazets, I. Lesanovsky, and J. Schmiedmayer, 
	\href{http://dx.doi.org/10.1103/PhysRevLett.105.265302}{Phys. Rev. Lett. {\bf 105}, 265302 (2010)}.
	
	\bibitem{FCFF11} 
	N. Fabbri, D. Clément, L. Fallani, C. Fort, and M. Inguscio, 
	\href{http://dx.doi.org/10.1103/PhysRevA.83.031604}{Phys. Rev. A {\bf 83}, 031604 (2011)}.
	
	\bibitem{DBAD12} 
	M. J. Davis, P. B. Blakie, A. H. van Amerongen, N. J. van Druten, and K. V. Kheruntsyan, 
	\href{http://dx.doi.org/10.1103/PhysRevA.85.031604}{Phys. Rev. A {\bf 85}, 031604 (2012)}.
	
	\bibitem{FPCF15} 
	N. Fabbri, M. Panfil, D. Clément, L. Fallani, M. Inguscio, C. Fort, and J.-S. Caux, 
	\href{http://dx.doi.org/10.1103/PhysRevA.91.043617}{Phys. Rev. A {\bf 91}, 043617 (2015)}.
	
	\bibitem{HSKDL08}
	G. Huber, F. Schmidt-Kaler, S. Deffner, and E. Lutz, \href{https://journals.aps.org/prl/abstract/10.1103/PhysRevLett.101.070403}{Phys. Rev. Lett. {\bf 101}, 070403 (2008)}. 
	
	\bibitem{DCHFGV13}
	R. Dorner, S. R. Clark, L. Heaney, R. Fazio, J. Goold, and V. Vedral, \href{https://journals.aps.org/prl/abstract/10.1103/PhysRevLett.110.230601}{Phys. Rev. Lett. {\bf 110}, 230601 (2013)}. 
	
	\bibitem{MDCP13}
	L. Mazzola, G. De Chiara, and M. Paternostro, \href{https://journals.aps.org/prl/abstract/10.1103/PhysRevLett.110.230602}{Phys. Rev. Lett. {\bf 110}, 230602 (2013)}. 
	
	\bibitem{CBKZH13}
	M. Campisi, R. Blattmann, S. Kohler, D. Zueco and P. Hänggi, \href{https://iopscience.iop.org/article/10.1088/1367-2630/15/10/105028/meta}{New J. Phys. {\bf 15} 105028 (2013)}. 
	
	\bibitem{Batalhaoetal}
	T. B. Batalh{\~a}o, A. M. Souza, L. Mazzola, R. Auccaise, R. S. Sarthour, I. S. Oliveira, J. Goold, G. De Chiara, M. Paternostro, and R. M. Serra, \href{https://journals.aps.org/prl/abstract/10.1103/PhysRevLett.113.140601}{Phys. Rev. Lett. {\bf 113}, 140601 (2014)}. 
	
	\bibitem{Anetal}
	S. An, J.N. Zhang, M. Um, D. Lv, Y. Lu, J. Zhang, Q.Y. Zhang, H.T. Quan and K. Kim, \href{https://www.nature.com/articles/nphys3197}{Nature Phys. {\bf 11}, 193-199 (2015)}. 
	
	\bibitem{olshanii-98} M. Olshanii, 
	\href{http://dx.doi.org/10.1103/PhysRevLett.81.938}{Phys. Rev. Lett. {\bf 81}, 938 (1998)}.
	
	\bibitem{iasm-98} S. Inouye, M. R. Andrews, J. Stenger, H.-J. Miesner, D. M. Stamper-Kurn, and W. Ketterle, 
	\href{http://dx.doi.org/10.1038/32354}{Nature {\bf 392}, 151 (1998)}.
	
	
	\bibitem{korepin_book}
	V. E. Korepin, N. M. Bogoliubov, and A. G. Izergin, Quantum Inverse Scattering Method and Correlation Functions (Cambridge University Press, Cambridge, UK, 1993).
	
	\bibitem{YaYa69} 
	C. N. Yang and C. P. Yang, 
	\href{http://dx.doi.org/10.1063/1.1664947}{J. Math. Phys. {\bf 10}, 1115 (1969)}.
	
	\bibitem{takahashi_book}
	M. Takahashi, Thermodynamics of One-Dimensional Solvable Models (Cambridge University Press, Cambridge, UK, 1999).
	
	
	\bibitem{GrRD10} 
	V. Gritsev, T. Rostunov, and E. Demler, 
	\href{http://dx.doi.org/10.1088/1742-5468/2010/05/P05012}{J. Stat. Mech. (2010) P05012}.
	
	\bibitem{KSCC13} 
	M. Kormos, A. Shashi, Y.-Z. Chou, J.-S. Caux, and A. Imambekov, 
	\href{http://dx.doi.org/10.1103/PhysRevB.88.205131}{Phys. Rev. B {\bf 88}, 205131 (2013)}.
	
	\bibitem{KoCC14} 
	M. Kormos, M. Collura, and P. Calabrese, 
	\href{http://dx.doi.org/10.1103/PhysRevA.89.013609}{Phys. Rev. A {\bf 89}, 013609 (2014)}.
	
	\bibitem{DeCa14} 
	J. De Nardis and J.-S. Caux, 
	\href{http://dx.doi.org/10.1088/1742-5468/2014/12/P12012}{J. Stat. Mech. (2014) P12012}.
	
	\bibitem{DePC15} 
	J. De Nardis, L. Piroli, and J.-S. Caux, 
	\href{http://dx.doi.org/10.1088/1751-8113/48/43/43FT01}{J. Phys. A: Math. Theor. {\bf 48}, 43FT01 (2015)}.
	
	\bibitem{PiCE16} 
	L. Piroli, P. Calabrese, and F. H. L. Essler, 
	\href{http://dx.doi.org/10.1103/PhysRevLett.116.070408}{Phys. Rev. Lett. {\bf 116}, 070408 (2016)};
	L. Piroli, P. Calabrese, and F. H. L. Essler, 
	\href{http://dx.doi.org/10.21468/SciPostPhys.1.1.001}{SciPost Phys. {\bf 1}, 001 (2016)}.
	
	\bibitem{ZWKG15} 
	J. C. Zill, T. M. Wright, K. V. Kheruntsyan, T. Gasenzer, and M. J. Davis, 
	\href{http://dx.doi.org/10.1103/PhysRevA.91.023611}{Phys. Rev. A {\bf 91}, 023611 (2015)};\\
	J. C. Zill, T. M. Wright, K. V. Kheruntsyan, T. Gasenzer, and M. J. Davis, 
	\href{http://dx.doi.org/10.1088/1367-2630/18/4/045010}{New J. Phys. {\bf 18}, 045010 (2016)};\\
	J. Zill, T. Wright, K. Kheruntsyan, T. Gasenzer, and M. Davis, 
	\href{http://dx.doi.org/10.21468/SciPostPhys.4.2.011}{SciPost Phys. {\bf 4}, 011 (2018)}.
	
	\bibitem{Broc14} 
	M. Brockmann, 
	\href{http://dx.doi.org/10.1088/1742-5468/2014/05/P05006}{J. Stat. Mech. (2014) P05006 }.
	
	\bibitem{BDWC14} 
	M. Brockmann, J. De Nardis, B. Wouters, and J.-S. Caux, 
	\href{http://dx.doi.org/10.1088/1751-8113/47/34/345003}{J. Phys. A: Math. Theor. {\bf 47}, 345003 (2014)}.
	
	\bibitem{BeSE14} 
	B. Bertini, D. Schuricht, and F. H. L. Essler, 
	\href{http://dx.doi.org/10.1088/1742-5468/2014/10/P10035}{J. Stat. Mech. (2014) P10035}.
	
	\bibitem{PiCa17} 
	L. Piroli and P. Calabrese, 
	\href{http://dx.doi.org/10.1103/PhysRevA.96.023611}{Phys. Rev. A {\bf 96}, 023611 (2017)}.
	
	
	\bibitem{KoPo12} 
	K. K. Kozlowski and B. Pozsgay, 
	\href{http://dx.doi.org/10.1088/1742-5468/2012/05/P05021}{J. Stat. Mech. (2012) P05021}.
	
	\bibitem{Pozs14} 
	B. Pozsgay, 
	\href{http://dx.doi.org/10.1088/1742-5468/2014/06/P06011}{J. Stat. Mech. (2014) P06011}.
	
	\bibitem{PiCa14} 
	L. Piroli and P. Calabrese, 
	\href{http://dx.doi.org/10.1088/1751-8113/47/38/385003}{J. Phys. A: Math. Theor. {\bf 47}, 385003 (2014)}.
	
	\bibitem{BNWC14} 
	M. Brockmann, J. D. Nardis, B. Wouters, and J.-S. Caux, 
	\href{http://dx.doi.org/10.1088/1751-8113/47/14/145003}{J. Phys. A: Math. Theor. {\bf 47}, 145003 (2014)}.
	
	\bibitem{LeKZ15} 
	M. de Leeuw, C. Kristjansen, and K. Zarembo, 
	\href{http://dx.doi.org/10.1007/JHEP08(2015)098}{JHEP 98 (2015)};\\
	I. Buhl-Mortensen, M. de Leeuw, C. Kristjansen, and K. Zarembo, 
	\href{http://dx.doi.org/10.1007/JHEP02(2016)052}{JHEP 52 (2016)};\\
	O. Foda and K. Zarembo, 
	\href{http://dx.doi.org/10.1088/1742-5468/2016/02/023107}{J. Stat. Mech. (2016) 023107}.
	
	\bibitem{LeKM16} 
	M. de Leeuw, C. Kristjansen, and S. Mori, 
	\href{http://dx.doi.org/10.1016/j.physletb.2016.10.044}{Phys. Lett. B {\bf 763}, 197 (2016)}.
	
	\bibitem{Pozs18} 
	B. Pozsgay, 
	\href{http://dx.doi.org/10.1088/1742-5468/aabbe1}{J. Stat. Mech. (2018) 053103}.
	
	\bibitem{LeKL18} 
	M. de Leeuw, C. Kristjansen, and G. Linardopoulos, 
	\href{http://dx.doi.org/10.1016/j.physletb.2018.03.083}{Phys. Lett. B {\bf 781}, 238 (2018)}.
	
	\bibitem{PiPV17} 
	L. Piroli, B. Pozsgay, and E. Vernier, 
	\href{http://dx.doi.org/10.1016/j.nuclphysb.2017.10.012}{Nucl. Phys. B {\bf 925}, 362 (2017)}.
	
	\bibitem{PoPV18} 
	B. Pozsgay, L. Piroli, and E. Vernier, 
	\href{http://arxiv.org/abs/1812.11094}{arXiv:1812.11094 (2018)}.
	
	
	
	\bibitem{DorGoold12}
	R. Dorner, J. Goold, C. Cormick, M. Paternostro, and V. Vedral \href{https://journals.aps.org/prl/abstract/10.1103/PhysRevLett.109.160601}{Phys. Rev. Lett. {\bf 109}, 160601 (2012)}. 
	
	\bibitem{PiPV17_II} 
	L. Piroli, B. Pozsgay, and E. Vernier, 
	\href{http://dx.doi.org/10.1088/1742-5468/aa5d1e}{J. Stat. Mech. (2017) 023106}.
	
	\bibitem{Klauser2011}
	A. Klauser and J.-S. Caux, \href{http://dx.doi.org/10.1103/PhysRevA.84.033604}{Phys. Rev. A {\bf 84}, 033604 (2011)}. 
	
	
	\bibitem{Takahashi1973}
	M. Takahashi, \href{https://academic.oup.com/ptp/article/50/5/1519/1839677}{Progr. Theor. Phys. {\bf 50}, 5 (1973)}. 
	
	\bibitem{BePi2018}
	B. Bertini and L. Piroli. \href{https://iopscience.iop.org/article/10.1088/1742-5468/aab04b/meta}{J. Stat. Mech. (2018) 033104}. 
	
	
	\bibitem{Gira60} 
	M. Girardeau, 
	\href{http://dx.doi.org/10.1063/1.1703687}{J. Math. Phys. {\bf 1}, 516 (1960)}.
	
	\bibitem{DoCa12} 
	P. L. Doussal and P. Calabrese, 
	\href{http://dx.doi.org/10.1088/1742-5468/2012/06/P06001}{J. Stat. Mech. (2012) P06001}.
	
	\bibitem{CaLe14} 
	P. Calabrese and P. Le Doussal, 
	\href{http://dx.doi.org/10.1088/1742-5468/2014/05/P05004}{J. Stat. Mech. (2014) P05004}.
	
	
	\bibitem{Zagier2012}
	D. Zagier, 
	\href{https://www.jstor.org/stable/pdf/23234630.pdf?seq=1#page_scan_tab_contents}{Ann. Math. {\bf 175}, 2 (2012)}.
	
	
	
	\bibitem{PiPV18} 
	L. Piroli, B. Pozsgay, and E. Vernier, 
	\href{http://dx.doi.org/10.1016/j.nuclphysb.2018.06.015}{Nucl. Phys. B {\bf 933}, 454 (2018)}.
	
	
	\bibitem{CaCR00} 
	L. D. Carr, C. W. Clark, and W. P. Reinhardt, 
	\href{http://dx.doi.org/10.1103/PhysRevA.62.063610}{Phys. Rev. A {\bf 62}, 063610 (2000)}.
	
	\bibitem{CaCR00_II} 
	L. D. Carr, C. W. Clark, and W. P. Reinhardt, 
	\href{http://dx.doi.org/10.1103/PhysRevA.62.063611}{Phys. Rev. A {\bf 62}, 063611 (2000)}.
	
	\bibitem{KaSU03} 
	R. Kanamoto, H. Saito, and M. Ueda, 
	\href{http://dx.doi.org/10.1103/PhysRevA.67.013608}{Phys. Rev. A {\bf 67}, 013608 (2003)}.
	
	\bibitem{FlFP16} 
	D. Flassig, A. Franca, and A. Pritzel, 
	\href{http://dx.doi.org/10.1103/PhysRevA.93.013627}{Phys. Rev. A {\bf 93}, 013627 (2016)}.
	
	\bibitem{PiCa16} 
	L. Piroli and P. Calabrese, 
	\href{http://dx.doi.org/10.1103/PhysRevA.94.053620}{Phys. Rev. A {\bf 94}, 053620 (2016)}.
	
	
	\bibitem{NIST}
	F. W. J. Olver, A. B. Olde Daalhuis, D. W. Lozier, B. I. Schneider, R. F. Boisvert, C. W. Clark, B. R. Miller, and B. V. Saunders, NIST Digital Library of Mathematical Functions, \href{http://dlmf.nist.gov/}{Release 1.0.22 of 2019-03-15}. 
	
\end{thebibliography}
\end{document}